\documentclass{jaa}
\usepackage[T1]{fontenc}
\usepackage{ae,aecompl}
\usepackage{graphicx}
\usepackage{amsmath}
\usepackage{amssymb}

\newcommand{\mkm}{$\mu$m}
\newcommand{\kms}{\,km\,s$^{-1}$}

\usepackage{longtable}

\begin{document}\sloppy

\title{High-resolution spectroscopy
of the variable hot post-AGB star LS~4331 (IRAS~17381-1616)}

\author{Natalia P. Ikonnikova\textsuperscript{1}, Mudumba Parthasarathy\textsuperscript{2,*}, Ivan A. Shaposhnikov\textsuperscript{1,3},
Swetlana Hubrig\textsuperscript{4} and Geetanjali Sarkar\textsuperscript{5}}
\affilOne{\textsuperscript{1} Lomonosov Moscow State University, Sternberg Astronomical Institute, 13 Universitetskij prospekt, Moscow 119234, Russia.\\}
\affilTwo{\textsuperscript{2} Indian Institute of Astrophysics, Bangalore 560034, India.\\}
\affilThree{\textsuperscript{3} Lomonosov Moscow State University, Faculty of Physics, 1 Leninskie Gory, Moscow 119991, Russia.\\}
\affilFour{\textsuperscript{4} Leibniz Institute for Astrophysics (AIP), Potsdam, D-14482, Germany.\\}
\affilFive{\textsuperscript{5} Department of Physics, Indian Institute of Technology,
Kanpur-208016, UP, India.}

\twocolumn[{

\maketitle

\corres{m-partha@hotmail.com}

\msinfo{18 January 2024}{18 January 2024}

\begin{abstract}

An analysis of high-resolution ($R\sim48\,000$) optical spectrum
of hot (B1Ibe) post-AGB star LS~4331 (IRAS~17381-1616) is
presented. The detailed identification of the observed absorption
and emission features in the wavelength range 3700--9200 \AA\ is
carried out for the first time. From non-LTE analysis of
absorption lines the atmospheric parameters and chemical
composition of the star are derived.  We estimate $T_{\rm
eff}=20~900\pm500$ K, $\log g=2.57\pm0.08$, $V_r=-51.7\pm0.8$
\kms, $\xi_{\rm t}=24\pm4$\, \kms ~and $v \sin i=30\pm5$ {\kms}. A
abundance analysis for C, N, O, Mg, Al, S and Si reveals that the
N and O abundance is close to solar while metal underabundances
relative to the solar value (i.e. [Mg/H] = --1.04 dex, [Al/H] =
--1.20 dex, [Si/H] = --0.46 dex) are found. We conclude that LS
4331 is a high galactic latitude metal-poor and carbon deficient
hot post-AGB star. The underabundance of carbon ([C/H]=--0.64 dex)
is similar to that found in other hot post-AGB stars and indicates
that the star's AGB phase of evolution was terminated before the
third dredge up. From the nebular emission lines the plasma
diagnostics are derived. The presence of nebular emission lines in
the spectrum of LS~4331 indicates that the photoionization of
circumstellar envelope has already started. The nebular parameters
and expansion velocity of the nebula is derived. Using the Gaia
DR3 distance the absolute luminosity of the star is derived and
the star's position on the post-AGB evolutionary tracks suggests
that its initial main sequence mass is about 1.2$M_{\odot}$. It is
also reported that fast irregular brightness variations with an
amplitude of up to 0.3 mag in $V$ band have been found in the
star, typical of hot post-AGB objects.

\end{abstract}


}]

\doinum{12.3456/s78910-011-012-3}
\artcitid{\#\#\#\#}
\volnum{000}
\year{0000}
\pgrange{1--}
\setcounter{page}{1}
\lp{1}

\section{INTRODUCTION}\label{sec:intro}

The late type (K, G, F) to hot (A, B, O) type post-AGB supergiants
form an evolutionary sequence in the transition region from the
tip of the AGB to early stages of planetary nebulae (Parthasarathy
1993, 1994).  The time taken for hot post-AGB supergiants to
evolve into a young planetary nebula is relatively very short and
it is possible to study this rapid phase of evolution in real time
in some cases. Analysis of  high resolution spectra of only very
few hot post-AGB stars has been carried out (Mooney {\em et al.}
2002, Klochkova {\em et al.} 2002, Sarkar {\em et al.} 2005, Mello
{\em et al.} 2012, Ikonnikova  {\em et al.} 2020 and references
therein). Some of the hot post-AGB stars are high galactic
latitudes and some are high velocity stars which suggests that
their progenitor stars are of low mass. There are hot post-AGB
stars in globular clusters. Therefore the study hot post-AGB stars
enable us to further understand the advanced stages of post-AGB
phase of evolution of low mass stars. Cool and hot post-AGB stars
were found among high galactic latitude stars and high velocity
stars (Parthasarathy \& Pottasch 1986, 1989, Parthasarathy {\em et
al.} 2020). Multiwavelength studies of these stars, their
circumstellar shells and chemical compositions proved that they
are indeed in the post-AGB stage of evolution. Only low and
intermediate mass stars go through the post-AGB stage of
evolution. Massive supergiants are not expected at high galactic
latitudes.  High galactic latitude supergiants were found to be in
post-AGB stage of evolution (Parthasarathy \& Pottasch 1986).
Low-mass stars during their post-AGB phase of evolution mimic the
spectra of  supergiants. Because they have thin extended
atmospheres around the CO core. Same is true with the high
velocity post-AGB stars. Chemical composition of some of the high
velocity supergiants and their IRAS flux distributions indicated
that they are in post-AGB stage of evolution  (for example,
V886~Her, LS IV --12 111, LS II +34 26 etc). Hence the progenitors
of these high galactic latitude and high velocity post-AGB stars
are low-mass stars.

In order to further understand this rapid phase of evolution,
chemical composition patterns, onset of the photoionization of the
shell, and late thermal pulse (LTP) objects it is important to
study many hot post-AGB stars. This is an effort in that
direction.

The star LS~4331 (Sp Type=OB+, Pmag=12.2 mag) from the Catalog of
Luminous stars in the Southern Milky Way by Stephenson \&
Sanduleak (1971) was found to be an infrared (IR) source
IRAS~17381-1616 with far-IR (IRAS) colours and far-IR flux
distribution similar to that of planetary nebulae and post-AGB
stars and hence it was classified as a high galactic latitude ($b=
+7.^{\circ}5$) hot post-AGB star (Parthasarathy {\em et al.}
2000). From a low resolution spectrum of this star (Parthasarathy
{\em et al.} 2000) found the spectral-type to be B1Ibe. Its IRAS
colours and IRAS flux distribution and characteristics of the
circumstellar dust shell and spectral type are similar to that of
post-AGB B-supergiant  LS II +34 26 (Parthasarathy 1993, 1994;
Garc\'{i}a-Lario {\em et al.} 1997). Su\'{a}rez {\em et al.}
(2006) classified the object as a planetary nebula. Cerrigone {\em
et al.} (2011) and references therein detected the radio continuum
indicating that it is in the very young planetary nebula stage.
Cerrigone {\em et al.} (2009) observed this star with the Spitzer
space telescope and detected silicate feature in the IR indicating
that the circumstellar envelope is Oxygen-rich. Some parameters of
LS~4331 are summarized in Table~\ref{table1}.


\begin{table*}
    \begin{center}
    \caption{Stellar and dust parameters of LS~4331 from the literature.}
    \label{table1}
    \begin{tabular}{ccc}
      \hline
      Quantity & Value & References \\
      \hline
      Position (J2000.0) & $\alpha = 17^{\mathrm{h}}41^{\mathrm{m}}00^{\mathrm{s}}$, $\delta = -16^{\circ}18^{\prime}12^{\prime\prime}$ (2000) & SIMBAD \\
      Gal. coord & $l = 010.255$, $b = +07.493$ & SIMBAD \\
      Proper motions (mas/yr) & $\mu_{\alpha} = -0.167\pm0.024$, $\mu_{\delta}=-9.421\pm0.015$ & G21 \\
      Parallaxes (mas) & $0.1086\pm0.0249$ & G21 \\
      Distance (pc) & 8256 (7439--9305)& BJ21 \\
      Spectral type & OB+, B1Ibe & SS71, P00 \\
      $T_{\rm eff}$ (K) & 24000 & C09 \\
      Magnitude &  13.57($B$) 13.22($V$) 13.22($R$) & UCAC4 \\
       & 13.36($g$) 13.07($r$) 13.11($i$)  & UCAC4 \\
       & 12.306($J$) 12.157($H$) 12.046($K_S$) & 2MASS \\
       & 13.084($G$) 13.251($BP$) 12.685($RP$) & G21\\
      $T_{\rm dust}$ (K) &
       250, 120 & C09 \\
      \hline
    \end{tabular}
    \end{center}
     References. SIMBAD: http://simbad.u-strasbg.fr/simbad/; SS71: Stephenson \& Sanduleak (1971); P00: Parthasarathy {\em et al.} (2000); C09: Cerrigone {\em et al.} (2009); UCAC4: Zachariah et al. (2013); G21: Gaia Collaboration {\em et al.} (2021); BJ21: Bailer-Jones {\em et al.} (2021)

\end{table*}


We have been carrying out detailed analysis of ESO FEROS high
resolution spectra of hot post-AGB stars and young planetary
nebulae (Otsuka {\em et al.} 2017, Arkhipova {\em et al.} 2018,
Ikonnikova {\em et al.} 2020, Herrero {\em et al.} 2020). There is
no detailed study of the spectrum of LS 4331. In this paper we
present a detailed analysis of the high resolution spectrum of LS
4331 in order to identify emission and absorption lines, and  to
understand its chemical composition and evolutionary stage. In
addition, we have studied the photometrical behavior of the star
and have discovered its rapid variability in brightness.

The paper is organized as follows: in Sect.~2 we describe the
observations and the data reduction; in Sect.~3  we present an
analysis of the main spectral features; the estimation of
atmospheric parameters and abundances are presented in Sect.~4. In
Sect.~5 we analyze the emission spectrum. Photometric observations
and their analysis are presented in Sect.~6. The evolutionary
status of the star is discussed in Sect.~7. In Sect.~8 we discuss
the results and in the last section we summarize our conclusions.

\section{Observations and data reduction}\label{sec2}

\subsection{High resolution spectroscopy}

The high-resolution optical spectrum of LS~4331 was obtained on
April 16, 2006 (MJD=53841.255) with the Fiberfed Extended Range
Optical Spectrograph (FEROS; Kaufer {\em et al.} 1999) on the
2.2-meter ESO/MPG telescope at La Silla Observatory, Chile
(Proposal ID.77.D-0478A, PI: M. Parthasarathy). Observations of
this star were carried out along with other hot post-AGB stars to
determine their chemical composition (see Otsuka {\em et al.}
2017, Arkhipova {\em et al.} 2018, Ikonnikova {\em et al.} 2020,
Herrero {\em et al.} 2020). The FEROS spectra have a resolution of
$R\sim48 000$ and a wavelength range from 3600 to 9200 \AA\ in 39
orders. The CCD detector EEV 2k$\times$4k with 15 \mkm\ pixel size
was used. The exposure time for each spectrum was 2700 s. The
signal-to-noise ratio was 100 per pixel at 5500 \AA.

The reduction process was performed using the FEROS standard
online reduction pipeline and the echelle spectra reduction
package ECHELLE in IRAF using a standard reduction manner
including bias subtraction, removing scattered light, detector
sensitivity correction, removing cosmic ray hits, airmass
extinction correction, flux density calibration, and an all
echelle order connection. Both reduced spectra were continuum
normalized, co-added, and cleaned of telluric lines with MOLECFIT
(Kausch {\em et al.} 2015).

\subsection{Photometric data}

\subsubsection{Multicolour photometry.}

\begin{figure}
\begin{center}
    \includegraphics[scale=0.6]{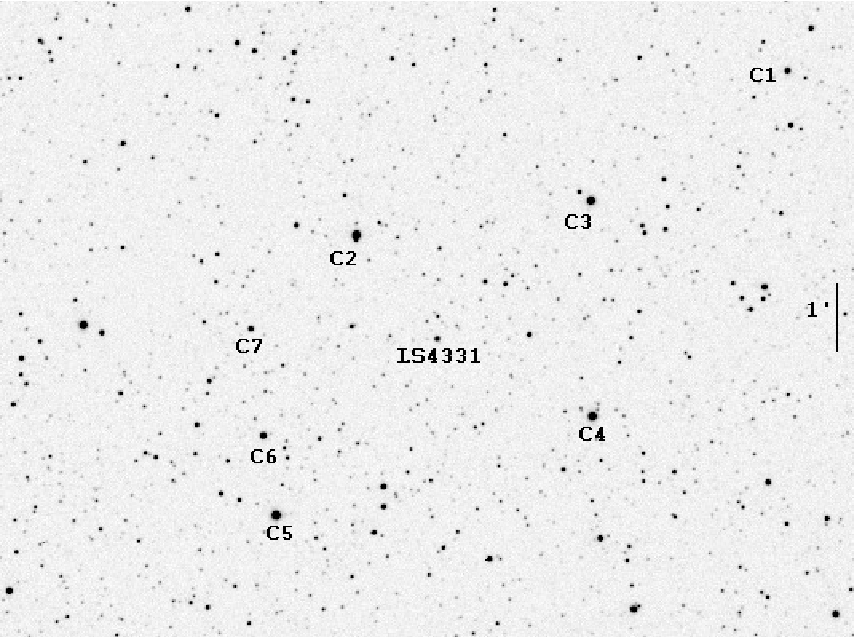}
    \caption{Finding chart for the field of LS~4331 in $R_C$. The abbreviations C1...C7 refer to the comparison stars mentioned in Table~\ref{table2}.}
    \label{fig1}
\end{center}
\end{figure}

Optical photometry for the star was obtained on the 60-cm
Ritchey-Cr\'{e}tien telescope (RC600) at the Caucasian mountain
observatory (CMO) of Moscow State University. The telescope is
equipped with a set of photometric filters and an Andor iKon-L CCD
(2048$\times$2048 pixels of 13.5~\mkm, the pixel scale is
$0\,.\!\!^{\prime\prime}67$~pixel$^{-1}$, the field of view is
$22^\prime\times22^\prime$). For a more detailed description of
the telescope and instrumentation we refer to Berdnikov {\em et
al.} (2020). We have observed LS~4331 for three seasons of
visibility in 2021--2023. A complete set of exposures for each
night consisted of 2--3 frames in each of the $BVR_{C}I_{C}$ and
$gri$ filters. The observations and data reduction (correction for
bias, dark, and flat fields) were performed using the MaxIm DL6
program\footnote{https://maxim-dl.software.informer.com/6.0/}.
Aperture photometry for each system was performed with the
AstroImageJ software package (Collins {\em et al.} 2017) using the
comparison stars noted in Table~\ref{table2}. The comparison
stars' magnitudes were adopted from Synthetic photometry generated
from the Gaia BP/RP mean
spectra\footnote{https://vizier.cds.unistra.fr/viz-bin/VizieR-3?-source=I/360/syntphot}.
Fig.~\ref{fig1} provides a finding chart for LS\,4331 and
comparison stars.


\begin{table*}
\caption{Comparison stars}
\footnotesize
 \label{table2}
 \begin{center}
\begin{tabular}{lccccccc}
\hline

Star&    $B$&     $V$&     $R_C$&      $I_C$&     $g$&    $r$&     $i$\\
\hline
C1&      13.114$\pm$0.015  &   12.555$\pm$0.008  &   12.206$\pm$0.006 &   11.848$\pm$0.005  &   12.770$\pm$0.010 &   12.388$\pm$0.007 &   12.250$\pm$0.005 \\
C2&      10.833$\pm$0.026  &   10.595$\pm$0.011  &   10.430$\pm$0.006 &   10.232$\pm$0.006  &   10.615$\pm$0.019 &   10.572$\pm$0.008 &   10.597$\pm$0.006 \\
C3&      11.513$\pm$0.025  &   11.041$\pm$0.010  &   10.742$\pm$0.008 &   10.432$\pm$0.008  &   11.200$\pm$0.014 &   10.913$\pm$0.009 &   10.822$\pm$0.008 \\
C4&      12.526$\pm$0.026  &   11.134$\pm$0.011  &   10.391$\pm$0.007 &    9.724$\pm$0.006  &   11.835$\pm$0.019 &   10.659$\pm$0.008 &   10.211$\pm$0.006 \\
C5&      12.070$\pm$0.022  &   10.844$\pm$0.010  &   10.174$\pm$0.006 &    9.550$\pm$0.005  &   11.442$\pm$0.016 &   10.429$\pm$0.007 &   10.024$\pm$0.006 \\
C6&      14.790$\pm$0.053  &   12.854$\pm$0.014  &   11.693$\pm$0.006 &   10.532$\pm$0.006  &   13.905$\pm$0.032 &   12.133$\pm$0.009 &   11.150$\pm$0.006 \\
C7&      14.625$\pm$0.027  &   13.278$\pm$0.012  &   12.518$\pm$0.007 &   11.797$\pm$0.006  &   13.961$\pm$0.019 &   12.803$\pm$0.008 &   12.297$\pm$0.006 \\
\hline
\end{tabular}
\end{center}
\end{table*}



We present the resulting photometry for LS~4331 in \ref{photom}
where for every night we list the mean time of observation and
magnitudes in each of photometric bands averaged over 2–3
frames. Our uncertainties defined as standard deviations for each
night and averaged over all nights are $\Delta B = 0.029$ mag,
$\Delta V = 0.013$ mag, $\Delta R_C = 0.009$ mag, $\Delta I_C =
0.014$ mag, $\Delta g = 0.021$ mag, $\Delta r = 0.010$ mag and
$\Delta i = 0.009$ mag.

\subsubsection{ASAS-SN data.}

To study the nature of the variability of the star, we also
investigated the much more numerous and wide-ranging observations
from All Sky Automated Survey for Super-Novae (ASAS-SN, Shappee
{\em et al.} 2014; Kochanek {\em et al.} 2017). The survey uses
several cameras on five different good astronomical sites to
acquire its data. The field of view of each camera is roughly 4.5
deg$^2$, the pixel scale is $8.^{\prime\prime}0$, and the image
FWHM is $\sim$2 pixels (Kochanek {\em et al.} 2017). The data are
available for download through the ASAS-SN web
page\footnote{https://asas-sn.osu.edu/}. Observations were
conducted using several small telescopes in two photometrical
bands: with a $V$ filter from the beginning of 2015 to the middle
of 2018, and with a $g$ filter from the beginning of 2018 until
now. ASAS-SN $V$-band data of LS~4331 span a time interval from
February 16, 2015 to September 23, 2018. ASAS-SN $g$-band data
(using the Sloan Digital Sky Survey (SDSS) $g$ filter) for LS~4331
became available from April 20, 2018 and continue until to October
22, 2023. We utilized images of good quality, excluding those
taken in poor weather  with $\rm FWHM >2$ pixels. We also removed
data with a large standard deviation (>0.03 mag). Finally, we were
left with 591 data points in the $V$ band and 2232 in the $g$ band
for analysis.

\section{Description of the high-resolution spectrum}\label{sec3}

\begin{figure*}
\begin{center}
    \includegraphics[scale=0.45]{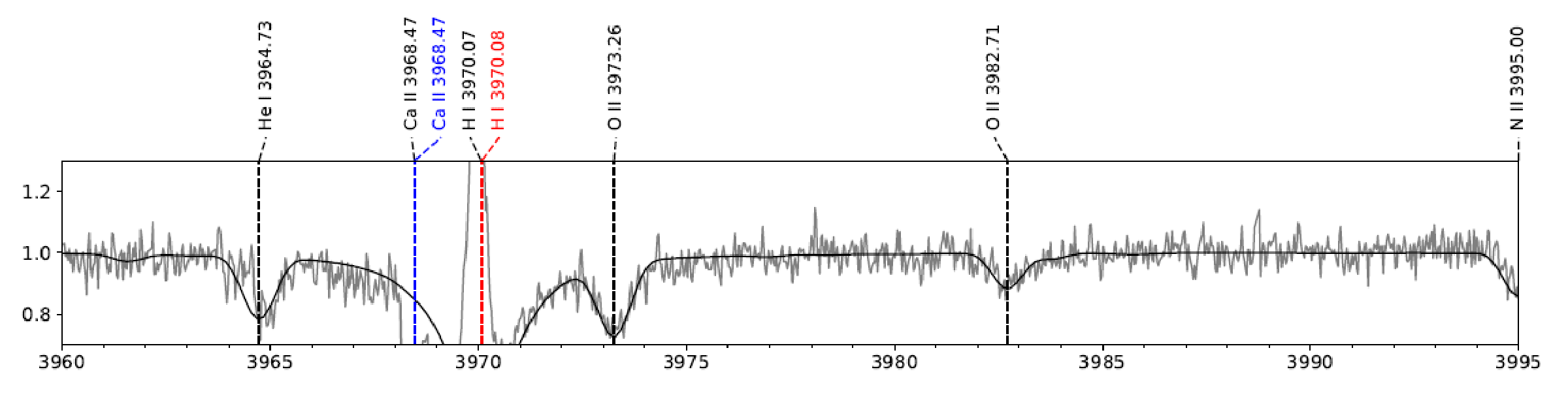}
    \caption{A fragment of the spectrum from the atlas.}
    \label{fig2}
\end{center}
\end{figure*}

The optical spectrum of LS~4331 shows stellar absorption lines,
nebula emission lines, and interstellar absorption features. Lines
identification are based on the National Institute of Standards
and Technology (NIST) Atomic Spectra
Database\footnote{https://www.nist.gov/pml/atomic-spectra-database}.
The detailed spectral atlas, which lists the observed and
synthetic spectrum of LS~4331, is presented at
\footnote{https://www.sai.msu.ru/research/2024/atlas-LS4331.pdf}.
Absorption features with the equivalent width (EW) of more than 35
m\AA\ are marked in black, identified emission lines are marked in
red, and features of the spectrum of interstellar origin are
marked in blue. The Fig.~\ref{fig2} shows a fragment of the
spectrum from the atlas.


\subsection{Photospheric absorption lines}

Absorption lines of neutral species including HI, HeI, and also
singly ionized species including NII, OII, SII, CII and MgII were
detected. Higher ionization is seen in AlIII, SiIII, SIII, CIII
and SiIV.

\subsection{Nebular emission lines}

The list of emission lines in LS~4331 is given in \ref{emlines} It
includes the measured and laboratory wavelength (in air),
equivalent width (EW), the heliocentric radial velocity, the name
of the element and the multiplet number to which the measured line
belongs.

\begin{figure}
\includegraphics[width=\columnwidth]{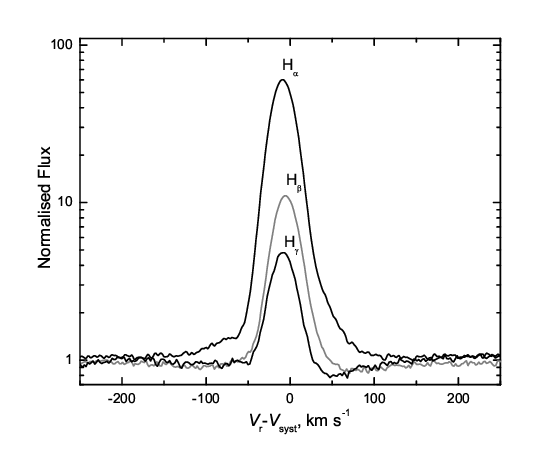}
\caption{Profiles of H$\alpha$, H$\beta$ and H$\gamma$
lines seen in the spectrum of LS~4331 on a velocity scale
relative to $V_{\rm sys} = -52$ \kms.} \label{fig3}
\end{figure}

The hydrogen Balmer and Pashen lines consist of the broad
photospheric absorption and nebular emission components. The
emission components of hydrogen lines do not show a P~Cyg profile.
The H$\alpha$ and H$\beta$ lines are fully filled in by emission
(Fig.~\ref{fig3}).

\begin{figure*}
    \includegraphics[scale=0.55]{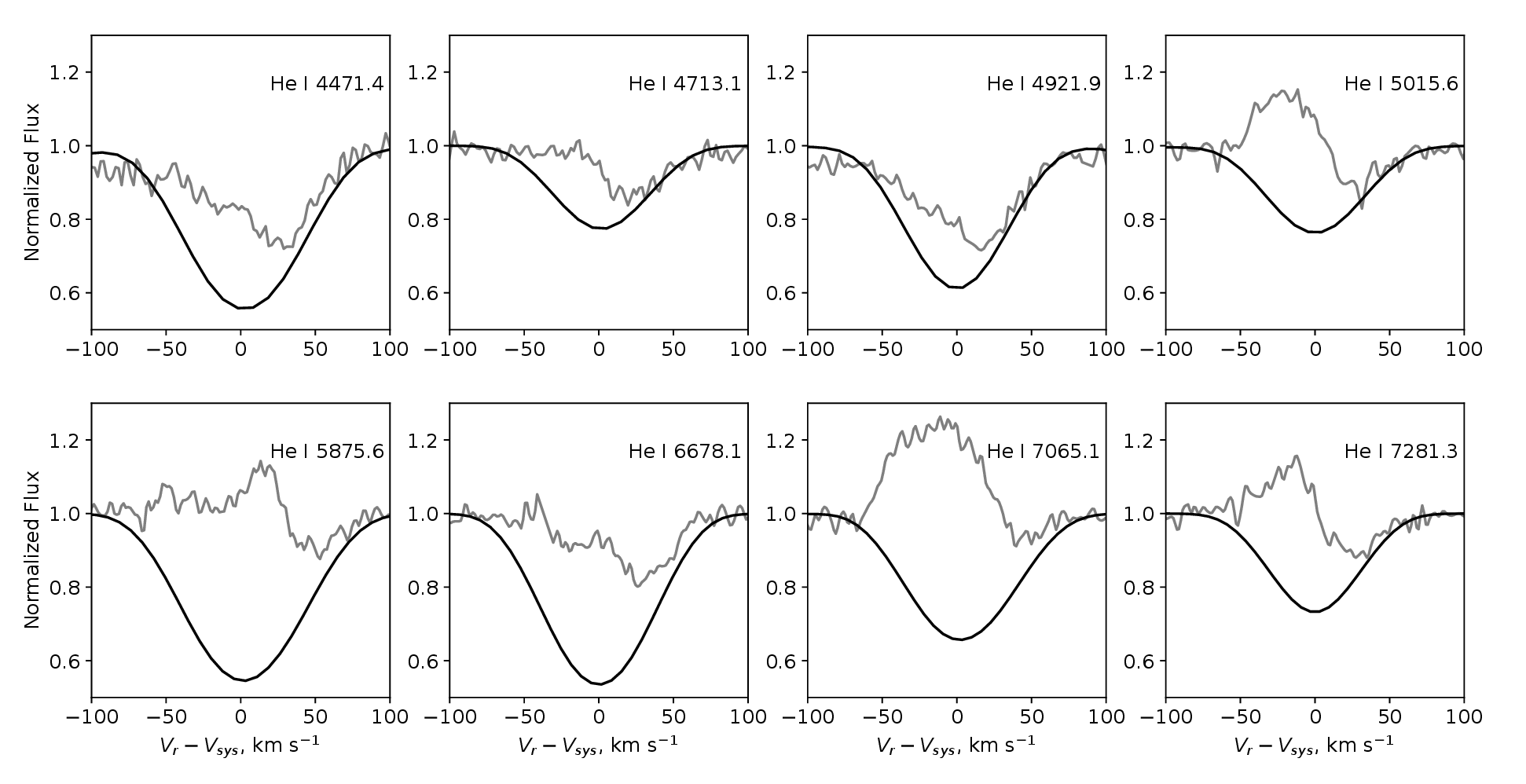}
    \caption{The profiles of selected HeI lines seen in the spectrum of LS~4331 on a velocity scale relative to $V_{sys} = -52$ \kms and compared to model spectra.}
    \label{fig4}
\end{figure*}

The helium lines are not included in \ref{emlines}  as they have
complicated multi-component profiles due to the emission lines
superposed on the corresponding absorption components.
Fig.~\ref{fig4} shows the profiles of selected HeI lines and
compared to model spectra (see next). The emission lines are
shifted to the left of the absorption lines, accordingly, they
have lower radial velocities.

The permitted emission lines, in addition to hydrogen and helium,
belong to the ions of SiII, CII, MgII, FeIII and also to the
nonionised atoms of OI and NI. In the red spectral region, the
permitted OI $\lambda$8446 triplet is the most remarkable emission
feature. The forbidden emission lines are from [FeII], [NiII]
$\lambda\lambda$7378, 7412, [NII] $\lambda\lambda$5755, 6548,
6584, [SII] $\lambda\lambda$6717, 6731, [CrII], [OI]
$\lambda\lambda$5577, 6300, 6363 as well as and [SIII]
$\lambda$9069.


\subsection{Interstellar features}

\begin{figure*}
    \centering
    \includegraphics[scale=0.80]{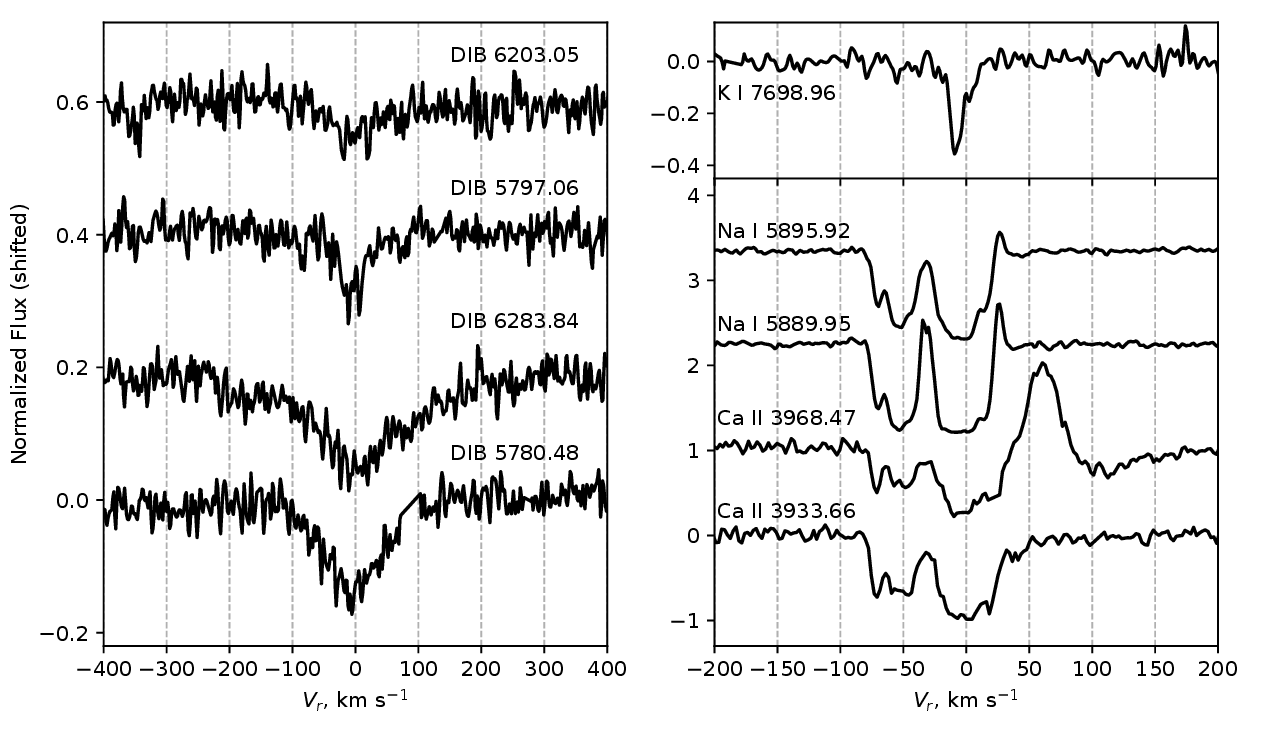}
    \caption{The profiles of interstellar absorption lines of CaII, NaI, and KI (the right panel) and four strongest DIBs (the left panel) in the spectrum of LS4331.}
    \label{fig5}
\end{figure*}

The spectrum of LS~4331 contains absorption features that have
interstellar origin. There are NaI doublet
($\lambda\lambda$5889.951, 5895.924), CaII  H and K lines
($\lambda\lambda$3968.469, 3933.663) and KI lines
($\lambda\lambda$7664.899, 7698.974). The H line of CaII is very
much blended with the strong stellar H$\epsilon$ line. The
selected interstellar spectral lines are depicted in the right
panel of Fig.~\ref{fig5}. The KI lines show a single and sharp
profiles with the radial velocities of --7.8$\pm$0.3 \kms ~and
--55.2$\pm$0.2 \kms. The NaI and CaII lines have broad
multi-component profiles. The main broad component contains
details that correspond to velocities of 0.5 \kms ~and --8.0 \kms.
The average velocity of the other broad component is
$V_r\approx-55$ \kms ~agrees with that of the short-wavelength
component of the KI line. As we have determined, the radial
velocity of the star is $-51.7\pm0.8$ \kms ~(see
Subsection~\ref{RV}), while the average velocity obtained from
emission lines of the circumstellar envelope corresponds to
$-60.3\pm4.7$ \kms ~(see Subsection~\ref{vel}). Therefore, it can
be deduced that the KI, CaII and NaI absorption lines with
velocities of circa --55  \kms ~are formed in the expanding
envelope around the central star.


In addition to the above-mentioned interstellar atomic lines, our
echelle spectra contain several quite strong diffuse interstellar
bands (DIBs): $\lambda\lambda$5781.48, 5798.06, 5850.81, 5889.95,
5895.92, 6196.98, 6204.05, 6285.04, 6380.62, 6614.82, 6994.13,
7116.31, 7120.71, and 7225.53, where the central wavelengths are
identified in accordance with Hobbs {\em et al.} (2008). The
profiles of several DIBs are shown in the left panel of
Fig.~\ref{fig4}. The average heliocentric velocity $V_r$ derived
from the DIBs interstellar absorption bands agrees with the
velocity of the NaI and CaII  main components: $V_r\approx-8.0$
\kms.

\section{Determination of the atmospheric parameters}\label{sec4}

The stellar parameters ($T_{\rm eff}$, $\log g$, $v \sin i$,
$\xi_{\rm t}$, $v\sin{i}$) and elemental abundances are determined
by fitting synthetic line profiles to the observed ones. The
synthetic profiles were calculated with SYNSPEC v.49 available at
\footnote{http://tlusty.oca.eu/Synspec49/synspec-frames-down.html},
using the BSTAR2006 grids generated with the code {\sc tlusty}
(Hubeny \& Lanz 1995), which assumes a plane-parallel atmosphere
in radiative, statistical non-local thermodynamic equilibrium
(non-LTE).

In cases where line profiles are distorted, for example by the
presence of emission components in the blue wing, the observed
profile was approximated by a synthetic profile through
minimization of $\chi^2$. Poor or distorted parts of the profile
were ignored.

The element abundance obtained from individual lines are averaged
with weights according to their uncertainties. Certain parameters
are interrelated, so by iteratively we derive a self-consistent
set of parameters.

\subsection{Radial velocity} \label{RV}

We selected 25 absorption lines, shapes of which are well fitted
by theoretical ones. We did not include lines with emission
features or with noticeably asymmetrical profiles. The wavelengths
shifts were found by fitting Gaussian profiles to both observed
and synthetic spectra in the same wavelength range. Average
measurements for each ion are presented in \ref{vr}  The weighted
average value along all lines gives the heliocentric velocity of
the star $V_r = -51.7\pm0.8$ \kms.

\subsection{Surface gravity $\log g$ and effective temperature $T_{\rm eff}$}

\begin{figure}
    \centering
    \includegraphics[scale=0.4]{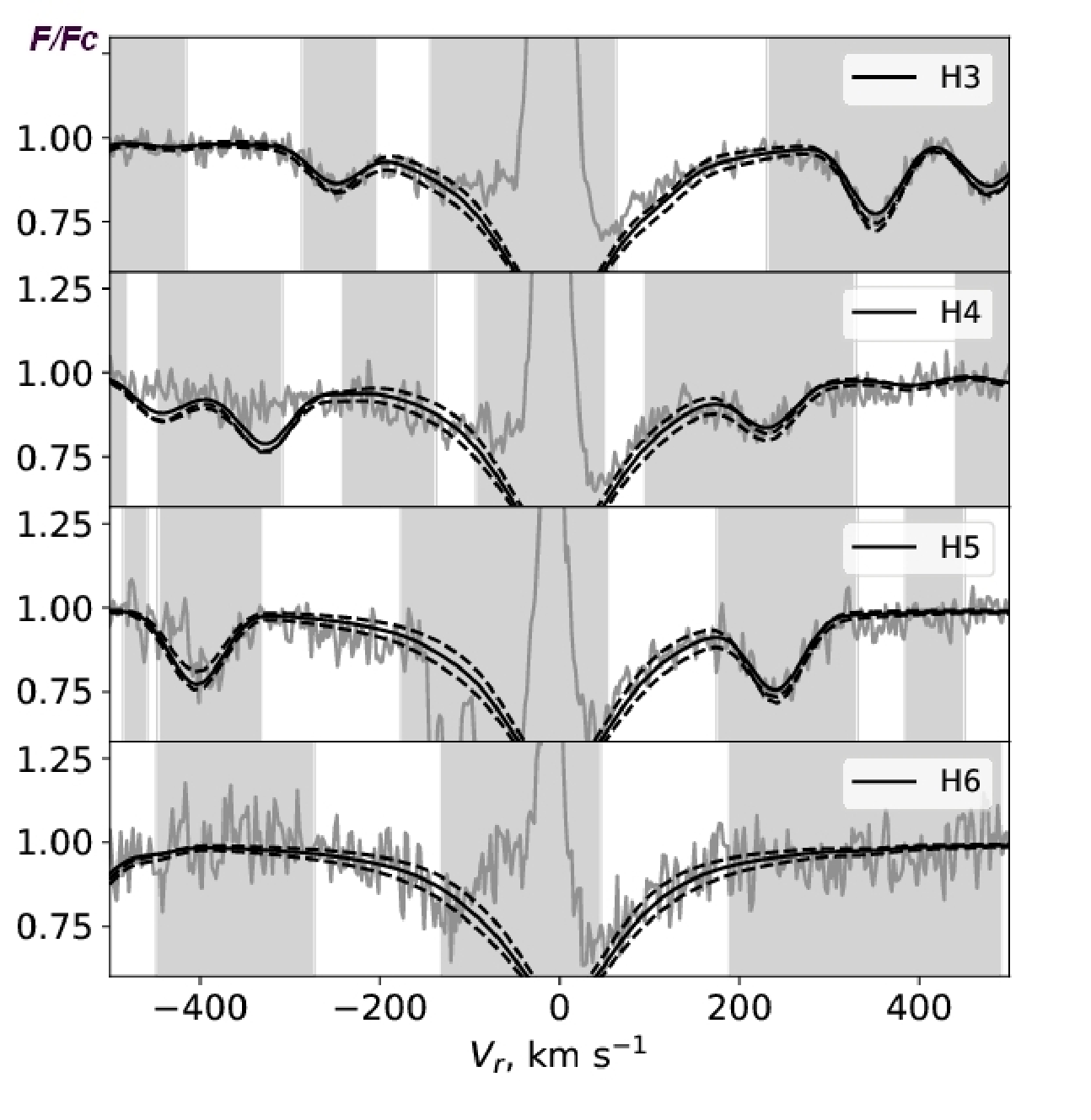}
    \caption{Wings of hydrogen lines. The shaded areas are ignored during
    the fit. The solid lines are for the best-fitting values of $\log g$.}
    \label{fig6}
\end{figure}

\begin{figure}
    \centering
    \includegraphics[scale=0.45]{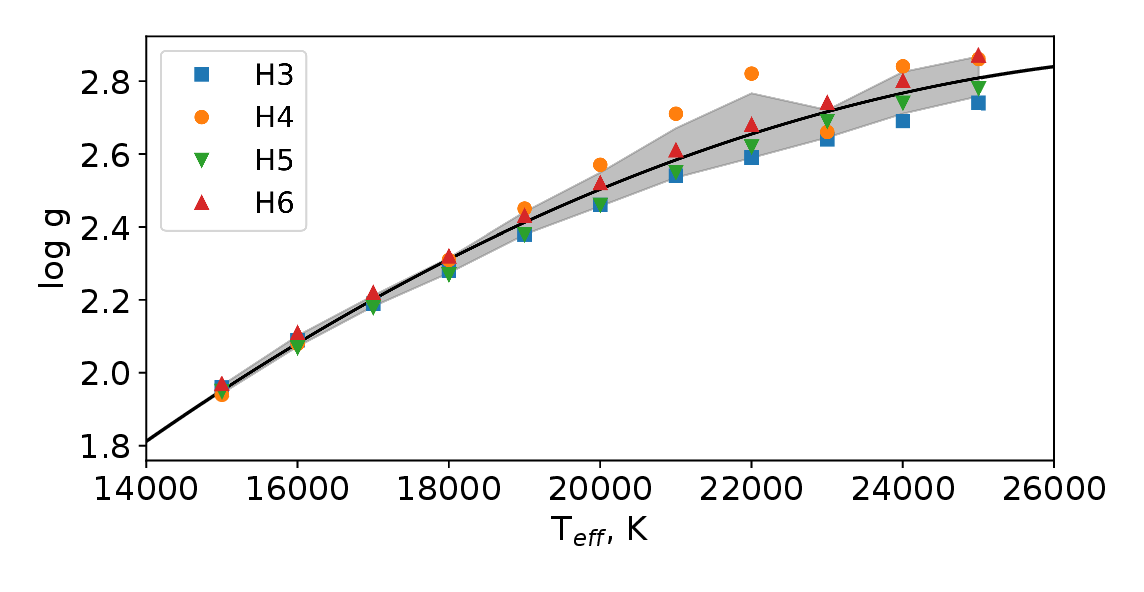}
    \caption{Parameters sequence estimated from hydrogen wings. Mean and sigma values were
    calculated for four lines at each point. The area between "mean-sigma" and "mean+sigma" is filled in gray.}
    \label{fig7}
\end{figure}

\begin{figure}
    \centering
    \includegraphics[scale=0.47]{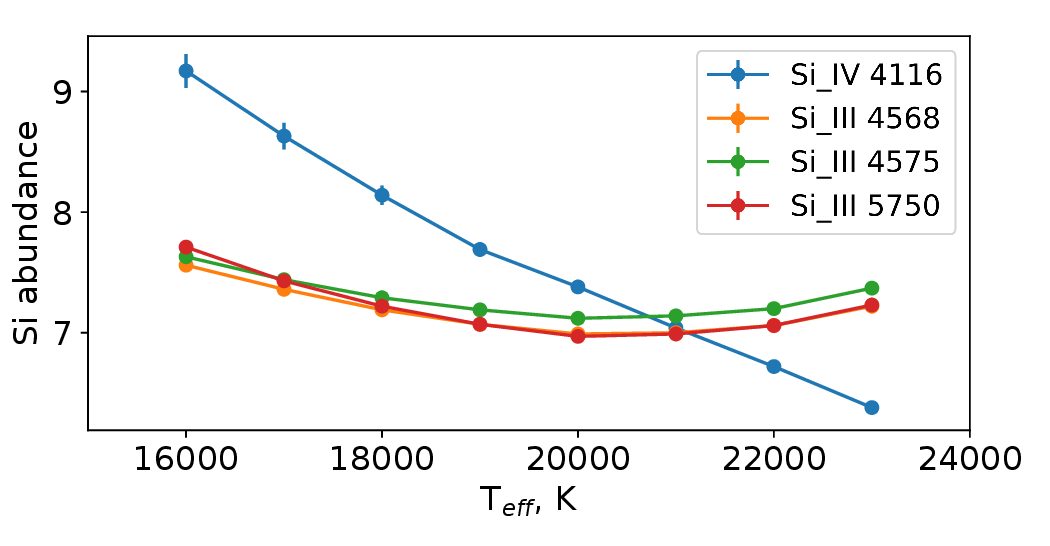}
    \caption{The silicon ionization balance.}
    \label{fig8}
\end{figure}

Surface gravity $\log g$ and effective temperature,
$T_\text{eff}$, were determined from the wings of hydrogen lines
and from the balance of ionization states of silicon (SiIII/SiIV).
The H$\alpha$ and H$\beta$ lines are filled in by emission
components (see Fig.~\ref{fig3}), so we used H$\gamma$, H$\delta$,
H$\epsilon$ and H6 (see Fig.~\ref{fig6}). By fitting the wings of
the hydrogen lines under different assumptions about the
temperature of the star, a sequence of $T_\text{eff}$-$\log g$
pairs is obtained, which can be described by the equation $\log g
= -1.01\pm0.38 + (2.65\pm0.41) \cdot 10^{-4}\cdot T_\text{eff} -
(4.5\pm1.0) \cdot 10^{-9} \cdot T_\text{eff}^2$ (Fig.~\ref{fig7}).
For these pairs of $\log g$ and $T_\text{eff}$, we select silicon
lines and adjust the $\epsilon$(Si) content for each individual
line to find the best fit. To do this, we selected only
non-blended lines: SiIII $\lambda\lambda$4567.8, 4574.8, 5739.7,
and SiIV $\lambda$4116.1. From Fig.~\ref{fig8}, it is clear that
the abundances measured from SiIII and SiIV lines are in agreement
with each other at $T_\text{eff} = 20900 \pm 500$ K and $\log g =
2.57 \pm 0.08$ and equal to $\log \epsilon(Si) = 7.04 \pm 0.03$.


\subsection{Microturbulence velocity}

\begin{figure*}
\centering

    \includegraphics[scale=0.75]{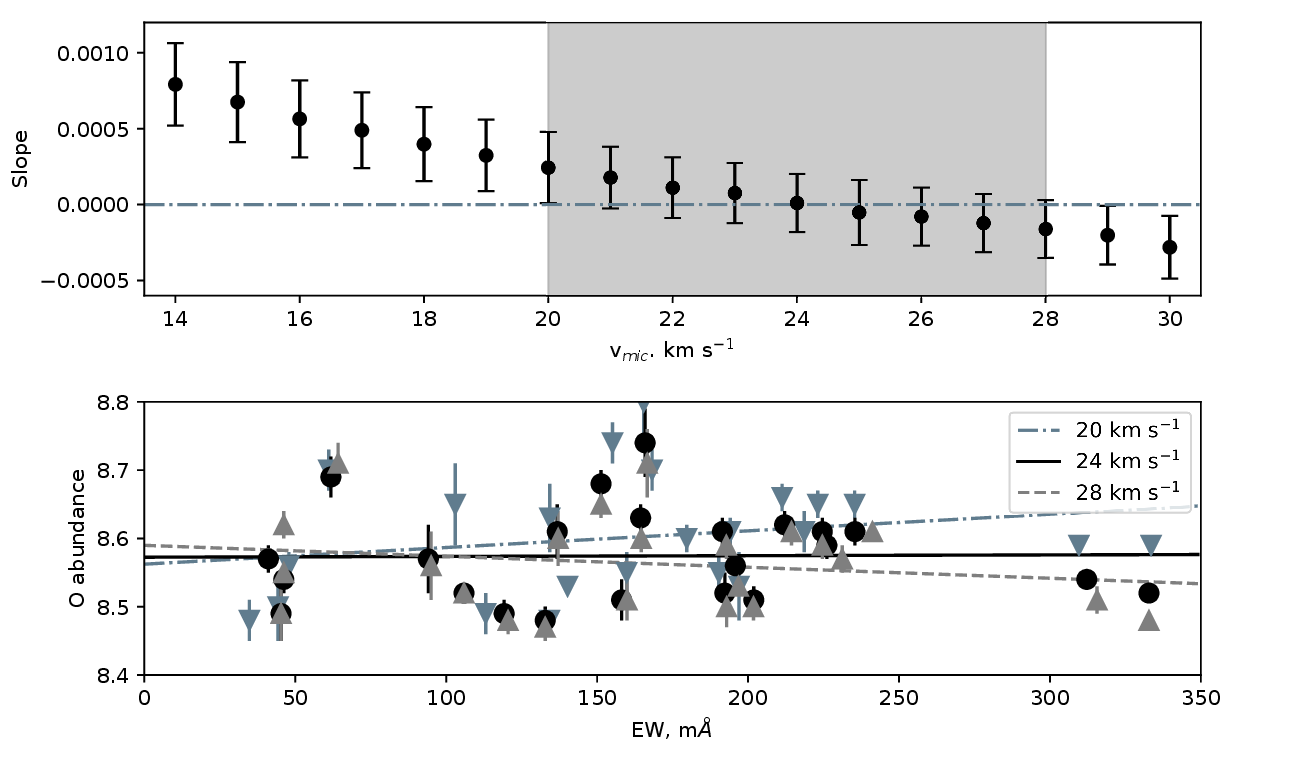}
    \caption{The abundance-equivalent width diagram to determine the microturbulence velocity.
    The top panel shows the dependence of the slope $\partial (\log \epsilon(O))/\partial (\rm EW)$
    on the estimated microturbulence velocity, the bottom panel shows a detailed "$\rm EW-\epsilon(O)$" diagram for three values of $\xi_{\rm t} = 20,24,28$ \kms.}
    \label{fig9}
\end{figure*}

The microturbulence velocity, $\xi_{\rm t}$, was derived from the
analysis of 23 non-blended OII lines. For each line, we adjust
oxygen abundance $\epsilon$(O) for wide range of trial values
$\xi_{\rm t}$. The obtained dependencies of $\epsilon$(O) on EW of
the lines are shown in Fig.~\ref{fig9}. The top panel shows the
dependence of the slope $\partial (\log \epsilon(O))/\partial (\rm
EW)$ on the estimated microturbulence velocity, the bottom panel
shows a detailed "$\rm EW-\epsilon(O)$" diagram for three values
of $\xi_{\rm t} = 20,24,28$ \kms. These slope dependencies were
calculated using the method of least squares, weighted by error.
Based on these data, we estimate the microturbulence velocity as
$\xi_{\rm t} = 24\pm4$ \kms. The weighted mean of $\log
\epsilon(O)$ = 8.57 with a standard error of 0.02 dex and a
standard deviation of 0.08 dex.


\subsection{Chemical abundances}

\begin{figure*}
\centering
    \includegraphics[scale=1.0]{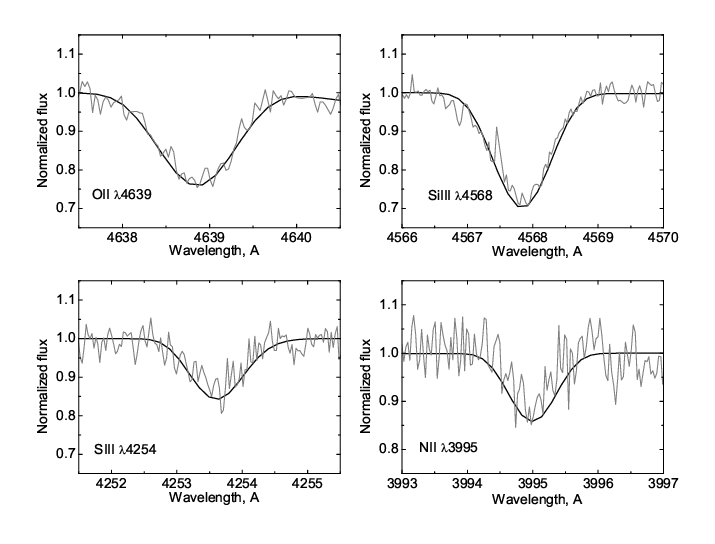}
    \caption{The non-LTE theoretical spectra (black lines) are overlaid with the observed profiles
    (grey lines) of the OII $\lambda$4639, SiIII $\lambda$4568, SIII $\lambda$4254, and NII $\lambda$3995 lines.}
    \label{fig10}
\end{figure*}

The elemental abundances were determined for the following
parameter set: $T_\text{eff}$ = 20900 K, $\log g$ = 2.57, and
$\xi_{\rm t}$ = 24 \kms. The non-LTE chemical abundances of 7
elements were obtained by matching the observed spectra with
theoretical profiles generated with the Synplot code. The results
for C, N, O, Mg, Al, Si and S were compiled into
Table~\ref{table3} along with their uncertainties.
Fig.~\ref{fig10} shows the observed and theoretical profiles of
some selected lines. We do not report the helium abundance in
Table~\ref{table3} as the obtained value of $\log\varepsilon({\rm
He})=10.61$ is unrealistic. The result is due to the fact that
helium lines have distorted profiles due to the influence of
emission components, which is visible on Fig.~\ref{fig4}.
Unfortunately, Ne and Fe lines in the spectrum LS~4331 are absent
or too weak to calculate the abundance. Absorption lines used for
measurement of the elemental abundances presented in \ref{vr}


\begin{table}
\caption{Element abundances determined for LS~4331 in $\log \varepsilon=12+\log(n_\text{X}/n_\text{H})$ and
[X/H]=$\log(n_\text{X}/n_\text{H})-\log(n_{\text{X}\odot}/n_{\text{H}\odot})$}
\begin{center}
\begin{tabular}{ccccccc}
      \hline
& $\log \varepsilon_{\odot}$   & $\log \varepsilon$ & [X/H]& $\sigma_x$&$\sigma_{\overline{x}}$ &N  \\

\hline
C  & 8.43 & 7.79 & -0.64 & 0.14 & 0.10 & 2  \\
N  & 7.83 & 7.45 & -0.38 & 0.30 & 0.12 & 6 \\
O  & 8.69 & 8.57 & -0.12 & 0.08 & 0.02 & 23 \\
Mg & 7.60 & 6.56 & -1.04 & -- & 0.04 & 1 \\
Al & 6.45 & 5.25 & -1.20 & -- & 0.05 & 1 \\
Si & 7.51 & 7.04 & -0.46 & 0.07 & 0.03 & 4 \\
S  & 7.12 & 6.31 & -0.81 & 0.06 & 0.05 & 2 \\
\hline
\end{tabular}
\end{center}


Solar values $\log \varepsilon_{\odot}$ are taken from
Asplund {\em et al.} (2009). $\sigma_x$ and $\sigma_{\overline{x}}$ are the standard deviation and the standard error of mean for $\log
\varepsilon.$
 \label{table3}
\end{table}


\begin{figure*}
\centering
    \includegraphics[scale=0.75]{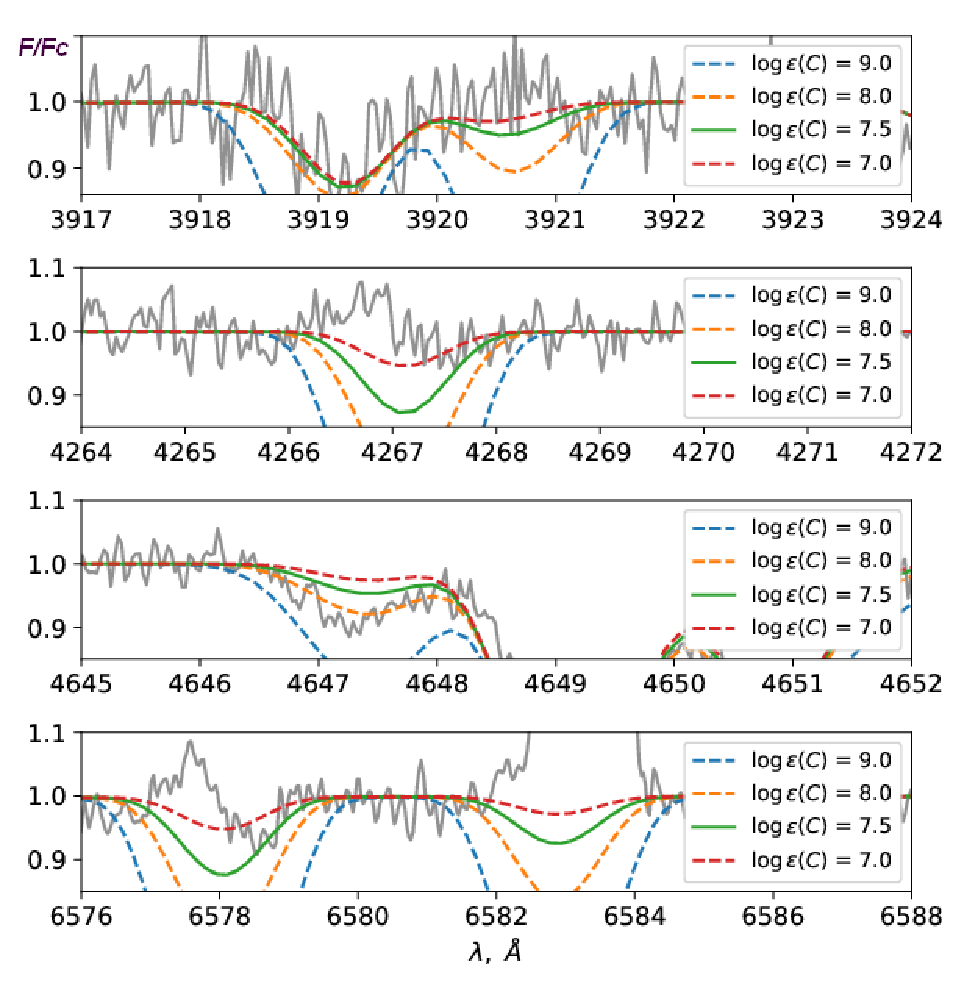}
    \caption{Observed CII ($\lambda \lambda$ 3919, 4267, 6578) and CIII ($\lambda 4647$) line profiles and synthetic ones for various $\log\varepsilon(C)$.}
    \label{fig11}
\end{figure*}

Determining the carbon abundance in LS 4331 is very difficult due
to the fact that CII lines, which are most commonly used for this
purpose ($\lambda \lambda$3921, 4227, and 4267), are extremely
weak. The CII lines at 5120--5152 \AA\ and 6780--6800 \AA, which
were used by Mello {\em et al.} (2012) to determine the carbon
abundance of a sample of hot post-AGB stars with parameters
similar to LS4331, are absent in the spectrum of LS4331 and cannot
be used by us. Fig.~\ref{fig11} shows observed profiles of some
CII and CIII lines and synthetic ones for various
$\log\varepsilon(C)$. The CII line $\lambda$3921 from the doublet
$\lambda\lambda$3919, 3921 is practically not visible. Fitting the
profile of the $\lambda$3719 line gives an abundance of
$7.25\pm0.25$; the abundance obtained from the EW measured over
the entire profile is $\log\varepsilon(C)=7.73\pm0.21$. In the
blue wing of the weak lines $\lambda$4267 and $\lambda$6578, there
are emission components, so the equivalent widths of these lines
cannot be measured. Fitting the red parts of the profiles gives an
abundance of $7.25\pm0.25$. The profile of CII $\lambda$6583 is
blended with the forbidden line NII $\lambda$6584. Fitting the
profile of the CIII $\lambda4647$ line gives a higher carbon
abundance value of $\sim8.0$. The equivalent width of this line,
EW=82 m\AA, results in an abundance value of $7.84\pm0.19$.
Therefore, to determine the abundance, we used only two lines, CII
$\lambda$3919 and CIII $\lambda$4647, and obtained an average
value of $\log\varepsilon({\rm C})=7.79$, which is significantly
lower than the solar abundance.


As is well known, the atmospheres of post-AGB stars are often
depleted in iron because it is part of the dust in their envelopes
and therefore cannot be used to determine the metallicity (Mathis
\& Lamers 1992). As an alternative, the initial metallicities of
stars can be estimated based on the abundance of sulfur
(Stasi\'{n}ska {\em et al.} 2006). As we determined, the sulfur
content in LS4331's atmosphere is estimated to be
[S/H]=-0.81$\pm$0.06, hence we can accept $Z=0.002$.

\subsection{Rotational velocity}

\begin{figure*}
    \centering
    \includegraphics[scale=0.80]{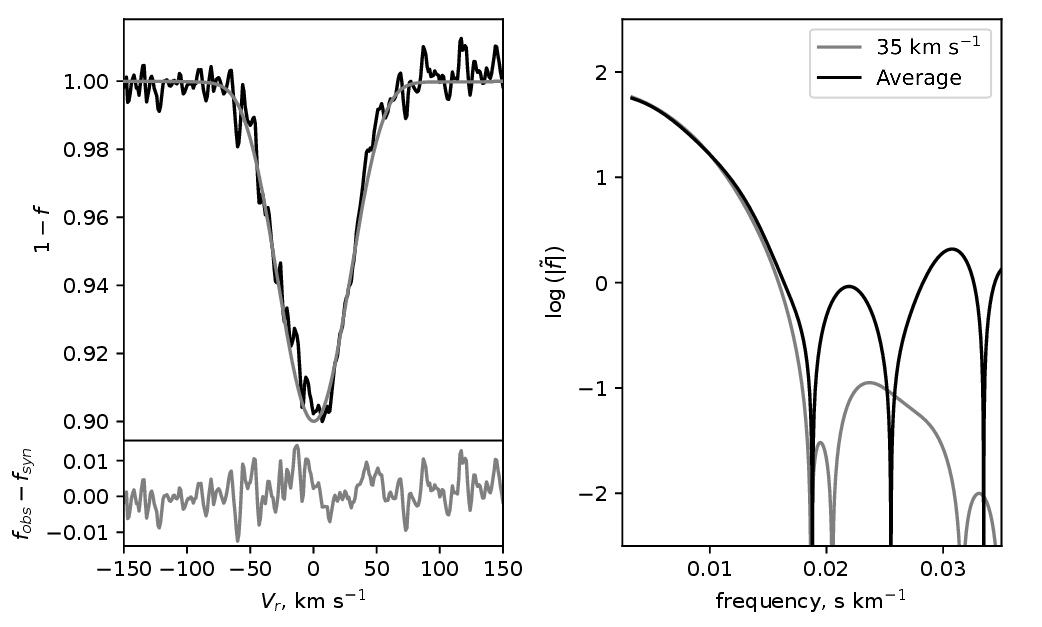}
    \caption{The average profile (the left panel) and corresponding Fourier transforms (the right panel).
    The black lines are for observations, the grey lines are for the synthetic spectrum for $v\sin\,i= 35$\,{\kms}.}
    \label{fig12}
\end{figure*}

The rotational velocity $v\sin{i}$ was determined by comparing the
averaged absorption profile with synthetic profiles using Fourier
transformation (Smith \& Gray 1976). To construct the average
profile we used five lines with complete profiles with no
distortions: SiIII $\lambda$4552, $\lambda$4568, $\lambda$4575,
$\lambda$5740 and OI $\lambda$4591. On Fig.~\ref{fig12} the
obtained average profile and the synthetic profile for the
rotation velocity 35 \kms are presented. The first zero of the
Fourier transform of the average profile corresponds to the first
zero of the Fourier transform for the synthetic spectrum
corresponding to the case of 35 km/s. To estimate the error in the
determination of $v\sin{i}$ we used synthetic observations
generated based on the optimal synthetic profile with $\sigma =
0.015$ by the Monte-Carlo method (a similar approach was used in
our work on LS 5112 (Ikonnikova {\em et al.} 2020). The final
value of $v\sin{i}$ we estimate as $35\pm7$ \kms.


\section{Analysis of the emission spectrum}\label{sec5}

Due to the rather low temperature of the central star, the
spectrum of the object still contains a quite limited set of lines
that are usually used for gas envelope diagnostics. Furthermore,
without a calibrated flux spectrum, it is not possible to obtain
reliable absolute fluxes of emission lines. However, using the
equivalent widths of emission lines from \ref{emlines} and the
distribution of stellar continuum fluxes for the above-mentioned
atmospheric parameters, we can reliably obtain ratios of fluxes in
emission lines.

\subsection{Nebular parameters}

From the available diagnostic line ratios, all densities and
temperatures were calculated with the code {\sc pyneb} (Luridiana
{\em et al.} 2015). {\sc pyneb} routine {\it getCrossTemden} was
used to determine simultaneously the temperature and the density
from the [NI], [NII], [OII] and [SII] lines, by building
diagnostic diagrams.

Densities can be determined from the [NI]
$\lambda\lambda$5198/5200, [SII] $\lambda\lambda$6717/6731 and
[OII] $\lambda\lambda$3726/3729 intensity ratios. Electron
temperatures can be obtained from the [NII]
$\lambda\lambda$(6548+6548)/5755, [SII]
$\lambda\lambda$(4069+4076)/(6717+6731) and [OII]
$\lambda\lambda$(3726+3729)/(7319+7330) intensity ratios.

\begin{figure}
    \centering
    \includegraphics[scale=0.5]{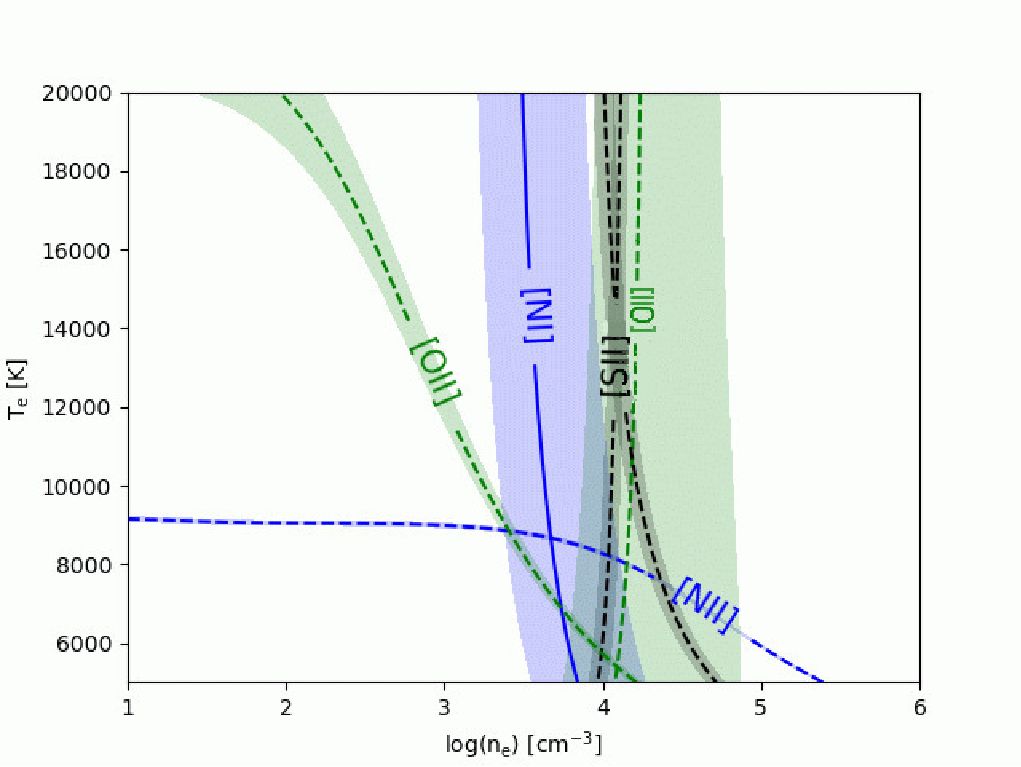}
    \caption{Diagnostic diagram for LS~4331, derived with {\sc pyneb}.}
    \label{fig13}
\end{figure}

Fig.~\ref{fig13} displays diagnostic diagram obtained using {\sc
pyneb}. In the $T_e$ range from 7000 to 20 000 K and $\log N_e$
(cm$^{-3}$) from 1 to 6, the curves have intersection. The nebular
line ratios for [NI] indicates that these lines are formed in a
partial ionized region of $N_e$ $\sim$ 3000 cm$^{-3}$, whereas
[SII] and [NII] intersection display electron density $N_e$ $\sim$
10000 cm$^{-3}$. The electron temperature of the nebula for this
region is estimated at $T_e$ = 8300-8800~K.


\subsection{Radial and expansion velocities of the gaseous nebula}\label{vel}

As a reminder, the stellar radial velocity, determined from
absorption lines, is equal to about --52 \kms
(subsection~\ref{RV}). The nebular radial velocity obtained from
emission lines is $-60.3\pm4.7$ \kms ~(\ref{emlines}). If the
shell had a spherical structure, these velocities should have been
the same. The difference suggests that the shape of the nebula
deviates from spherical. In support of this conclusion, we can
point to the finding by Cerrigone {\em et al.} (2008) that the
nebula at radio maps at 8.4 GHz observed in high angular
resolution has a morphology that resembles a bipolar structure.

To calculate the expansion velocity of the gaseous envelope, we
used eight of the strongest and most reliably measured forbidden
emission lines. We used the following formula: $V_{\rm exp} = 1/2
\sqrt{V_{\rm FWHM}^2 - V_{\rm instr}^2}$, where $V_{\rm FWHM}$ is
the velocity corresponding to the FWHM and $V_{\rm instr}$ = 6
\kms ~is the instrumental broadening. The adopted $V_{\rm exp}$
for each ion are given in Table~\ref{table4}.


\begin{table}
    \centering
    \small
    \caption{Expansion velocities obtained from emission lines.}
    \begin{tabular}{ccccc}
      \hline
$\lambda_{obs}$,& $\lambda_{lab}$,& ion & FWHM,& $V_{exp}$,\\
\AA & \AA & &\AA & \kms  \\
\hline
4286.55  &  4287.39  &  [Fe II](7F)  &  0.46   & 15.8 \\
6299.04  &  6300.30  &  [O I](1F)    &  0.72   & 16.8 \\
6546.75  &  6548.05  &  [N II](1F)   &  0.58   & 12.9 \\
6582.13  &  6583.45  &  [N II](1F)   &  0.58   & 12.9 \\
6715.12  &  6716.44  &  [S II](2F)   &  0.65   & 14.2 \\
6729.48  &  6730.82  &  [S II](2F)   &  0.65   & 14.2 \\
7318.64  &  7319.99  &  [O II]       &  0.70   & 14.1 \\
9067.12  &  9068.80  &  [S III]      &  0.63   & 10.0 \\
      \hline
    \end{tabular}
    \label{table4}
\end{table}


Note the difference in the expansion velocity between the less and
more ionised species. The highest value was obtained for the [OI]
line (16.8 \kms), the lowest for the [SIII] line (10.0 \kms), and
intermediate values for [NII], [OII], [SII] and [Fe II] (12.9 --
14.2 \kms). This can be explained by the fact that the ionized
shell possesses a layered structure, with more ionized species
located closer to the central star. The structure tends to display
an expansion velocity gradient, with larger velocities directed
towards the external zones, where less-ionized species are found.

\section{Photometric analysis}\label{sec6}

\begin{figure*}

    \includegraphics[scale=1.5]{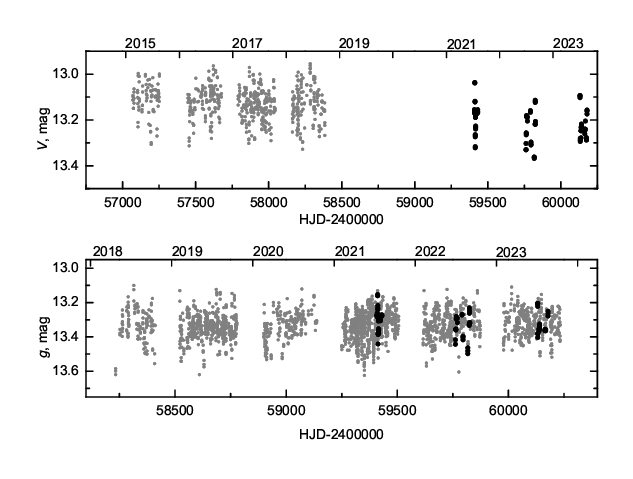}
    \caption{The light curves of ASAS-SN (grey circles) and RC600 (black circles) data in $V$-band (top panel) and $g$-band (bottom panel).}
    \label{fig14}
\end{figure*}

As seen in Fig.~\ref{fig14}, which shows the light curves in both
the $V$ and $g$ bands, based on the combined ASAS-SN and RC600
data, the object is clearly variable with peak-to-peak range of
brightness variations of about 0.3 mag and 0.4 mag, respectively.
The brightness of the star varies significantly from night to
night. The behavior of the brightness in the other bands is
similar to that of the $V$-band. We conducted a periodogram
analysis of the ASAS-SN light curves using the period-finding
program Period04 (Lenz \& Breger 2005), but did not detect any
periodic components in the period range of 0.3-50 days in the full
dataset. We also did not detect any periodicity when analyzing the
data for each season individually.

\begin{figure}
\centering
\includegraphics[scale=1.45]{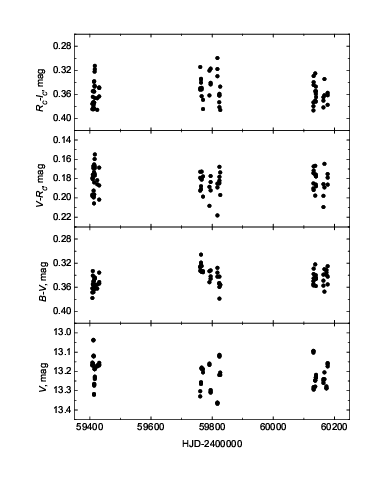}
    \caption{The light and colour curves of  RC600  data in 2021-2023.}
    \label{fig15}
\end{figure}

\begin{figure}
    \centering
    \includegraphics[scale=1.2]{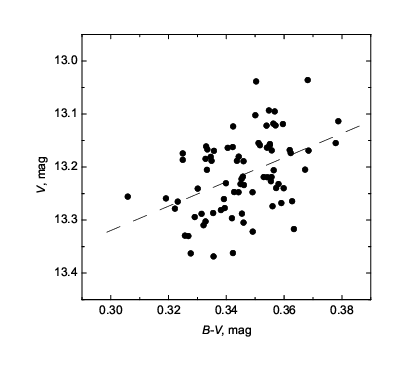}
    \caption{Plots showing the change in colour with change in brightness. The object shows a trend
    of becoming bluer when it fainter. The dashed line represents a linear fit to the data.}
    \label{fig16}
\end{figure}

The multicolour photometry obtained with the RC600 telescope
allowed us to track the change in colour alongside the brightness.
Fig.~\ref{fig15} shows the $V$-band light curve  and the $B-V$,
$V-R_C$, $R_C-I_C$ colour curves of the star for three seasons in
2021-2023. As can be seen on Fig.~\ref{fig15}, colour indices
change both from season to season and within each season.
Fig.~\ref{fig16} illustrates the dependencies of the colour
indices $B-V$ and $R_C-I_C$ on the brightness. It can be seen
that, as the brightness decreases, $B-V$ decreases, while
$R_C-I_C$ increases, making the star look redder.

Similar variability in brightness and colour $B-V$ has been
observed in our previous research on the study of hot post-AGB
stars. For example, IRAS~19200+3457 (Arkhipova {\em et al.} 2004),
V886 Her (Arkhipova {\em et al.} 2007), IRAS 01005+7910, IRAS
22023+5249 and IRAS 22495+5134 (Arkhipova {\em et al.} 2013)
exhibit rapid variability with a period of a few days and an
amplitude of 0.2 to 0.4 mag in the $V$ band. So far, no periodic
component has been detected in the variability of brightness for
any of the objects in our studies. Like LS4331, the other objects
mentioned exhibit a decrease in their $B-V$ colour index when the
object dims.

Hot post-AGB stars can be variable in brightness due to various
processes that affect their surface and atmosphere, such as
pulsation, mass loss, and evolution. Evolutionary changes occur
over time scales of several decades, as in the case of V886 Her
(Arkhipova {\em et al.} 2007). Pulsations can also occur, but so
far for all of the above objects, a periodic component in the
change in brightness has not been identified.  Additionally, hot
post-AGB stars do not exhibit an increase in their $B-V$ colour
index when their brightness decreases, as is characteristic of
pulsating stars. For example, cooler post-AGB stars of spectral
type F-G-K, which pulsate, become redder when they fade,
indicating a decrease in temperature due to the pulsations (e.g.,
Hrivnak {\em et al.} 2015; Ikonnikova {\em et al.} 2023).

Thus, to explain variability with a characteristic time of a
fraction of a day or several days and an amplitude of the order of
0.2-0.4 mag, a variable stellar wind remains. According to
evolutionary calculations (Miller Bertolami 2016), for a hot
post-AGB star with a temperature in the range of 21000 K, an
estimated rate of mass loss on the order of $10^{-8}M_{\odot} $
per year is expected.


\section{Mass and luminosity}\label{LM}

Doubts about LS~4331 belonging to the class of post-AGB stars are
not valid. In addition to the main characteristics of this class
of objects such as the presence of an excess of emission in the
far-infrared region and the shape of the spectral energy
distribution, in this work we defined some parameters of the star
and its gaseous envelope characteristic of stars in the post-AGB
stage of evolution.

\begin{figure}
    \centering
    \includegraphics[scale=0.9]{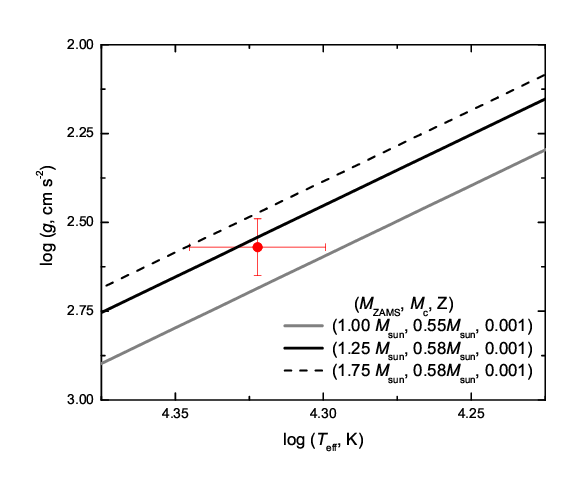}
    \caption{Location of LS~4331 (the red circle) in the $ T_{\rm eff}$-$\log g$ plane with $Z$ = 0.001
    single-star evolutionary tracks of Miller Bertolami (2016).}
    \label{fig17}
\end{figure}

To determine the stellar mass, we compared the obtained values of
effective temperature and surface gravity with the evolutionary
tracks for H-rich post-AGB stars that have been presented by
(Miller Bertolami 2016). The closest to the obtained point is the
track with the parameters: the initial mass of the progenitor
$M_{\rm ZAMS} = 1.25M_{\odot}$ and the current mass $M_{\rm c} =
0.58M_{\odot}$ for initial metallicity $Z = 0.001$
(Fig.~\ref{fig17}). The star with $T_{\rm eff}=21000$ K falls on
the horizontal part of the post-AGB evolutionary track on the
Hertzsprung-Russell diagram with $\log(L/L_{\odot}) \sim$ 3.9.

On the one hand, the luminosity of the star can be determined by
knowing the distance, the interstellar extinction, and the
apparent brightness. The average brightness of $V$ in our data is
$13.22\pm0.07$ mag, while the colour index $B-V=0.34\pm0.02$. For
a supergiant of effective temperature $T_{\rm eff} \approx21000$ K
($\log T_{\rm eff} = 4.32$) the normal colour index $(B-V)_0$ is
--0.2 based on the calibration of Flower (1996). Then the colour
excess $E(B-V)$ is $0.54\pm0.02$. The star in the DR3 catalog has
the parallax of $0.1086 \pm 0.0249$ (Gaia Collaboration {\em et
al.} 2021). The median of the photogeometric  distance derived
from this parallax is $d=8256_{-819}^{+1049}$ pc (Bailer-Jones
{\em et al.} 2021).  According to the known formula for absolute
stellar magnitude $M_V = V + 5 - 5 \log d - A_V$, where $A_V$ is
the extinction in $V$ band, calculated as $A_V = 3.1 E(B-V)$, we
find $M_V\approx-3.04$. For a star of effective temperature
$T_{\rm eff} \approx 21000$ K according to a calibration of Flower
(1996) the bolometric correction is estimated to be $BC=-2.0$, and
thus the absolute bolometric magnitude is given by $M_{\rm bol} =
M_V + BC\approx-5.04$ and $\log(L/L_{\odot})$ can be calculated as
$\log(L/L_{\odot}) = (M_{\rm bol\odot} - M_{\rm
bol})/2.5\approx3.90$. This evaluation is in close agreement with
that obtained by comparing with evolutionary models.

It should be taken into account that both luminosity estimates are
burdened with significant errors. In the case of an estimate
obtained by comparing the parameters of the star with evolutionary
models, the evolutionary track we have chosen does not fully agree
with the characteristics of the star, since its metallicity
$Z=0.002$ is slightly higher than that accepted for the model
($Z=0.001$). Calculations for $Z=0.002$ are not presented in
Miller Bertolami (2016). In addition, the parameters of the star
($\log T_{\rm eff}$ and $\log g$), which were compared to
evolutionary tracks, also have some errors associated with them.
Also, each of the quantities included in the formula for
calculating the absolute magnitude (primarily distance) has
significant errors.

\section{DISCUSSION}\label{disc}

Our analysis of the high-resolution optical spectrum of  LS~4331
together with our detailed line identifications  confirms that
LS4331 is a evolving hot post-AGB star. The chemical composition
of LS 4331 reported in this study clearly indicates that it is
metal-poor and O-rich similar to that found in some of the hot
post-AGB stars (Moehler \& Heber 1998, Ryans {\em et al.} 2003 and
references therein) and indicates that its AGB phase of evolution
may have been terminated before the third dredge up. We concluded
that LS4331 is metal-poor based on the following facts: firstly,
the iron lines are extremely weak, and we were unable to measure
their equivalent widths to determine the chemical composition.
Additionally, the abundance of elements such as Mg, Al, and S is
significantly lower than solar (Table 3). The carbon lines are
weak and the carbon abundance is determined with uncertainty,
while the oxygen abundance is measured reliably and is close to
the solar value. Thus, our study confirms the conclusion of
Cerrigone {\em et al.} (2009) that LS4331 is an O-rich object,
similar to some other hot post-AGB objects, such as
IRAS~17074-1845, IRAS~17203-1534, IRAS~17460-3114, and
IRAS~18062+2410 (Cerrigone {\em et al.} 2009), for the first three
of which Mello {\em et al.} (2012), and for IRAS~18062+2410 Ryans
{\em et al.} (2003) determined the chemical composition and
confirmed that in their stellar atmospheres C/O<1. Examples of hot
O-rich post-AGB objects also include IRAS~19590-1249 (Ryans {\em
et al.} 2003, Mello {\em et al.} 2012), PG 1323-086 and
PG~1704+222 (Moehler \& Heber 1998), and IRAS~18379–1707
(Ikonnikova {\em et al.} 2020).

To determine the iron abundance in the stellar atmosphere,
high-resolution spectral observations in the ultraviolet (UV)
region would be useful. Such work has been done for post-AGB stars
in globular clusters: ROA 5701 in $\omega$ Centauri and Barnard~29
in M~13. For these stars, the UV observations are from the Goddard
high-resolution spectrograph on the Hubble Space Telescope (HST).
The UV data provide additional Fe abundance estimates from Fe III
absorption lines in the 1875--1900 \AA\ wavelength region
(Thompson {\em et al.} 2007).

The analysis of the high-resolution spectrum did not reveal any
signs of the presence of a second component in the system.
However, for a more conclusive conclusion about the absence of
binarity, a long-term spectral monitoring is necessary, which
would allow to construct a radial velocity curve.

The effective temperature of the star ($T_{\rm eff}\approx21
000$~K) is high enough to ionize the gaseous envelope surrounding
the star. Analysis of the emission spectrum of the star allowed us
to estimate the parameters of the gas envelope.

The discovery of rapid irregular photometric variability in
LS~4331 is an important result of our work. Currently, about 15
objects with similar photometric characteristics are known  (for
example, V1853~Cyg (Turner \& Drilling 1984, Arkhipova {\em et
al.} 2001), V886~Her (Arkhipova {\em et al.} 2007),
IRAS~07171+1823 (Arkhipova {\em et al.} 2006), IRAS~19306+1407
(Hrivnak {\em et al.} 2020) etc), which allows us to consider this
to be a common property of hot post-AGB stars. The nature of this
variability is still not well understood, and further photometric
and spectroscopic observations are necessary to study the physical
processes occurring in the atmospheres of these stars.

\section{CONCLUSIONS}\label{concl}

We have presented a non-LTE analysis of high resolution  optical
spectrum, and photometric variability of LS~4331, a hot post-AGB
star whose spectral properties and photometric behavior had not
been investigated before in detail. Our conclusions can be
summarized as follows:

1) We identified the spectrum in the wavelength range of 3600-9500
Å for the first time. The spectrum contains numerous absorption
lines belonging to the star and emission lines from the
surrounding gas envelope.

2) We determined the parameters of the star: effective temperature
$T_{\rm eff}=20900\pm500$ K, surface gravity $\log g=2.57\pm0.08$,
radial velocity $V_r=-51.7\pm0.8$ \kms, and micro-turbulent
velocity $\xi_{\rm t}=24\pm4$ \kms.

3) We determined the composition of 7 elements ([C/H]=--0.64 dex,
[N/H]=--0.38 dex, [O/H]=--0.12 dex, [Mg/H]=--1.04, [Al/H] = --1.20
dex, [Si/H] = --0.46 dex, [S/H] = --0.81 dex) and confirmed the
conclusion that the star is rich in oxygen and poor in carbon and
metals.

4) The star already has a high enough temperature to ionise its
envelope, and we estimated parameters $T_e=5000$ K and
$N_e=3000-5000$ $\rm cm^{-3}$ for it.

5) By comparing with theoretical evolutionary tracks, we
determined the initial  mass $M_{\rm ZAMS} \sim 1.25M_{\odot}$ and
the current mass $M_{\rm c} \sim 0.58M_{\odot}$ of the star. The
estimated luminosity in the range of $L=5600-7300$ $L_\odot$ is
obtained both by comparison with theoretical evolutionary
sequences and by using distance measurements from the Gaia DR3.

6) Analysis of the light curves obtained in the ASAS-SN survey and
by us on the RC600 telescope showed that the star experiences
irregular brightness fluctuations with a full amplitude of about
0.3 mag in the $V$ band and a typical period of about one to
several days. It is hypothesized that the variability in
brightness is related to the variable stellar wind. We estimated
the colour excess to be $E(B-V)=0.54\pm0.02$.

\section*{Acknowledgement}

This research has used the SIMBAD data base, operated at CDS,
Strasbourg, France, and SAO/NASA Astrophysics Data System.
We are thankful to the referee very much for useful comments and constructive
suggestions.

\appendix

\newpage

\onecolumn
\section{Multicolour photometry of LS~4331} \label{photom}
\small
\begin{center}
\begin{longtable}{cccccccc}
  \caption{Multicolour photometry of LS~4331.} \label{tab:photom}\\

\hline

HJD    &    $B$&     $V$&     $R_C$&      $I_C$&     $g$&    $r$&     $i$\\
\hline

 \hline

  \endfirsthead

   \multicolumn{2}{l}{continued Table \ref{tab:photom}}\\
   \hline
HJD    &    $B$&     $V$&     $R_C$&      $I_C$&     $g$&    $r$&     $i$\\
   \hline
   \endhead

2459408.325 & 13.530 & 13.168 & 12.970 & 12.594 & 13.272 & 12.967 & 12.982 \\
2459408.329 & 13.533 & 13.155 & 12.975 & - & 13.276 & 12.975 & 12.959 \\
2459408.332 & 13.538 & 13.169 & 12.973 & 12.588 & 13.279 & 12.973 & 12.978 \\
2459409.293 & 13.494 & 13.161 & 12.982 & 12.628 & 13.277 & 12.982 & 13.012 \\
2459409.296 & 13.514 & 13.159 & 12.983 & 12.608 & 13.261 & 12.983 & 13.001 \\
2459409.299 & 13.504 & 13.164 & 12.983 & 12.619 & 13.283 & 12.983 & 13.005 \\
2459411.277 & 13.404 & 13.036 & 12.869 & 12.496 & 13.161 & 12.869 & 12.887 \\
2459411.280 & 13.389 & 13.039 & 12.869 & 12.488 & 13.154 & 12.869 & 12.896 \\
2459412.262 & 13.474 & 13.118 & 12.926 & 12.549 & 13.218 & 12.926 & 12.922 \\
2459412.265 & 13.476 & 13.122 & 12.923 & 12.569 & 13.232 & 12.923 & 12.923 \\
2459412.268 & 13.478 & 13.121 & 12.922 & 12.538 & 13.241 & 12.922 & 12.927 \\
2459413.301 & 13.671 & 13.322 & 13.116 & 12.777 & 13.441 & 13.116 & 13.161 \\
2459413.305 & 13.680 & 13.317 & 13.144 & 12.760 & 13.384 & 13.144 & -- \\
2459414.295 & 13.630 & 13.274 & 13.077 & 12.722 & 13.385 & 13.077 & 13.119 \\
2459414.298 & 13.627 & 13.265 & 13.081 & 12.742 & 13.377 & 13.080 & 13.121 \\
2459414.301 & 13.627 & 13.268 & 13.102 & 12.729 & 13.376 & 13.102 & 13.118 \\
2459415.296 & 13.535 & 13.189 & 13.022 & 12.675 & 13.301 & 13.022 & -- \\
2459415.300 & 13.525 & 13.180 & 13.020 & 12.698 & 13.302 & 13.020 & 13.058 \\
2459415.303 & 13.532 & 13.188 & 13.014 & 12.668 & 13.308 & 13.014 & -- \\
2459416.294 & 13.582 & 13.227 & 13.072 & 12.706 & 13.378 & 13.072 & -- \\
2459416.297 & 13.600 & 13.240 & 13.071 & 12.758 & 13.362 & 13.071 & -- \\
2459416.300 & 13.577 & 13.232 & 13.055 & 12.737 & 13.353 & 13.055 & -- \\
2459424.254 & 13.524 & 13.169 & 12.987 & 12.602 & 13.307 & 12.987 & 13.002 \\
2459424.257 & 13.536 & 13.174 & 12.988 & 12.621 & 13.291 & 12.988 & 12.990 \\
2459430.272 & 13.518 & 13.164 & 12.962 & 12.612 & 13.277 & 12.962 & 13.012 \\
2459430.275 & 13.506 & 13.155 & 12.986 & 12.623 & 13.272 & 12.986 & 13.017 \\
2459430.278 & 13.505 & 13.169 & 12.982 & 12.634 & 13.277 & 12.982 & 13.001 \\
2459761.329 & 13.636 & 13.303 & 13.130 & 12.815 & 13.417 & 13.130 & 13.171 \\
2459761.332 & 13.657 & 13.330 & 13.151 & 12.799 & 13.417 & 13.151 & 13.175 \\
2459761.335 & 13.655 & 13.330 & 13.137 & 12.788 & 13.443 & 13.137 & 13.183 \\
2459763.377 & 13.562 & 13.256 & 13.076 & 12.741 & 13.360 & 13.076 & 13.135 \\
2459763.380 & 13.578 & 13.259 & 13.071 & 12.731 & 13.354 & 13.071 & 13.113 \\
2459763.383 & 13.588 & 13.265 & 13.074 & 12.740 & 13.357 & 13.074 & 13.140 \\
2459765.327 & 13.517 & 13.185 & 13.005 & 12.656 & 13.288 & 13.005 & 13.031 \\
2459765.330 & 13.511 & 13.187 & 13.006 & 12.654 & 13.295 & 13.006 & 13.027 \\
2459765.333 & 13.515 & 13.181 & 13.008 & 12.645 & 13.282 & 13.008 & 13.029 \\
2459770.318 & 13.539 & 13.205 & 13.006 & 12.638 & 13.290 & 13.006 & 13.044 \\
2459770.322 & 13.523 & 13.188 & 13.010 & 12.626 & 13.317 & 13.010 & 13.037 \\
2459791.282 & 13.511 & 13.159 & 12.970 & 12.609 & 13.273 & 12.970 & 12.997 \\
2459791.285 & 13.500 & 13.167 & 12.959 & 12.638 & 13.268 & 12.959 & 12.985 \\
2459795.249 & 13.651 & 13.305 & 13.121 & 12.778 & 13.396 & 13.121 & 13.166 \\
2459795.252 & 13.642 & 13.310 & 13.117 & 12.800 & 13.417 & 13.118 & 13.154 \\
2459795.255 & 13.638 & 13.297 & 13.119 & 12.779 & 13.403 & 13.119 & 13.161 \\
2459817.248 & 13.690 & 13.363 & 13.174 & 12.856 & 13.463 & 13.174 & 13.233 \\
2459817.249 & 13.704 & 13.368 & 13.150 & 12.851 & 13.481 & 13.150 & 13.242 \\
2459817.250 & 13.704 & 13.362 & 13.177 & 12.847 & 13.498 & 13.177 & 13.245 \\
2459823.216 & 13.573 & 13.219 & 13.034 & 12.661 & 13.333 & 13.034 & 13.045 \\
2459823.219 & 13.572 & 13.219 & 13.035 & 12.673 & 13.326 & 13.035 & 13.045 \\
2459824.199 & 13.466 & 13.123 & 12.945 & 12.584 & 13.262 & 12.945 & 12.949 \\
2459824.200 & 13.492 & 13.114 & 12.931 & 12.573 & 13.230 & 12.931 & 12.950 \\
2459824.201 & 13.479 & 13.119 & 12.951 & 12.570 & 13.246 & 12.951 & 12.952 \\
2459826.197 & 13.574 & 13.219 & 13.022 & 12.675 & 13.321 & 13.022 & -- \\
2459826.201 & 13.563 & 13.206 & 13.033 & 12.647 & 13.315 & 13.032 & -- \\
2460131.330 & 13.448 & 13.093 & 12.919 & 12.539 & 13.221 & 12.919 & 12.928 \\
2460131.333 & 13.452 & 13.102 & 12.912 & 12.539 & 13.205 & 12.912 & 12.918 \\
2460131.336 & 13.452 & 13.095 & 12.910 & 12.524 & 13.206 & 12.910 & 12.931 \\
2460132.369 & 13.633 & 13.288 & 13.116 & 12.772 & 13.380 & 13.116 & 13.157 \\
2460132.372 & 13.619 & 13.281 & 13.114 & 12.774 & 13.383 & 13.114 & 13.146 \\
2460132.375 & 13.623 & 13.294 & 13.102 & 12.773 & 13.403 & 13.102 & 13.159 \\
2460137.310 & 13.592 & 13.248 & 13.081 & 12.756 & 13.353 & 13.081 & 13.148 \\
2460137.313 & 13.601 & 13.278 & 13.081 & 12.726 & 13.357 & 13.081 & 13.141 \\
2460137.316 & 13.590 & 13.247 & 13.071 & 12.701 & 13.367 & 13.071 & 13.12 \\
2460139.322 & 13.590 & 13.232 & 13.056 & 12.697 & 13.355 & 13.056 & 13.071 \\
2460139.326 & 13.580 & 13.234 & 13.056 & 12.685 & 13.342 & 13.056 & 13.092 \\
2460139.329 & 13.570 & 13.230 & 13.044 & 12.697 & 13.347 & 13.044 & 13.064 \\
2460140.306 & 13.568 & 13.223 & 13.035 & 12.680 & 13.325 & 13.035 & 13.076 \\
2460140.309 & 13.564 & 13.218 & 13.040 & 12.676 & 13.336 & 13.040 & 13.064 \\
2460164.248 & 13.599 & 13.260 & 13.051 & 12.669 & 13.355 & 13.051 & 13.066 \\
2460164.251 & 13.597 & 13.240 & 13.054 & 12.684 & 13.364 & 13.054 & 13.063 \\
2460164.254 & 13.597 & 13.248 & 13.050 & 12.685 & 13.363 & 13.050 & 13.083 \\
2460167.256 & 13.572 & 13.205 & 13.040 & 12.679 & 13.364 & 13.040 & 13.106 \\
2460167.259 & 13.571 & 13.241 & 13.052 & 12.717 & 13.356 & 13.052 & 13.095 \\
2460174.239 & 13.619 & 13.288 & -- & 12.763 & -- & -- & -- \\
2460174.242 & 13.622 & 13.287 & -- & 12.749 & -- & -- & -- \\
2460174.246 & 13.617 & 13.277 & -- & 12.738 & - & 13.014 & -- \\
2460178.232 & 13.504 & 13.162 & 12.987 & 12.610 & 13.265 & 12.987 & 13.032 \\
2460178.235 & 13.512 & 13.157 & 12.978 & 12.617 & 13.278 & 12.978 & 13.005 \\
2460178.238 & 13.499 & 13.174 & 12.988 & 12.631 & 13.250 & 12.988 & 13.015 \\
\hline

\end{longtable}
\end{center}

\onecolumn
\section{Emission lines in the spectrum of LS~4331}\label{emlines}
\small
\begin{center}
\begin{longtable}{cccccc}
  \caption{Emission lines in the spectrum of LS~4331.}
  \label{tab:emlines}\\

\hline

   $\lambda_{\text{obs.}}$, \AA& $\lambda_\text{lab.}$, \AA & Identification &  $EW$, m\AA &  $\sigma_{EW}$, m\AA & $V_\text{r}$, {\kms} \\

  \hline

  \endfirsthead

   \multicolumn{2}{l}{continued Table \ref{tab:emlines}}\\

   \hline
   $\lambda_{\text{obs.}}$, \AA& $\lambda_\text{lab.}$, \AA & Identification &  $EW$, m\AA &  $\sigma_{EW}$, m\AA & $V_\text{r}$, {\kms} \\

   \hline
   \endhead

3728.07  &  3728.82  & [O II] & 1497  & 137 & -60.30 \\
3725.30  &  3726.03  & [O II] & 3529  & 149 & -58.74 \\
3728.07  &  3728.82  & [O II] & 1497  & 137 & -60.30 \\
4067.79  &  4068.59  &  [S II]       & 408  & 47 &  -58.9  \\
4243.19  &  4243.98  &  [Fe II](21F) & 83   & 13 &  -55.8  \\
4286.55  &  4287.39  &  [Fe II](7F)  & 226  & 11 &  -58.7  \\
4339.59  &  4340.47  &  H I          & 1996 & 20 &  -60.4  \\
4358.46  &  4359.34  &  [Fe II](7F)  & 185  & 17 &  -60.5  \\
4367.40  &  4368.13  &  O I(5)       & 152  & 21 &  -50.0  \\
4412.87  &  4413.78  &  [Fe II](6F)  & 203  & 23 &  -61.7  \\
4813.53  &  4814.55  &  [Fe II](20F) & 119  & 21 &  -63.5  \\
4860.35  &  4861.28  &  H I          & 5693 & 27 &  -57.4  \\
5039.98  &  5041.03  &  Si II(5)     & 134  & 12 &  -62.4  \\
5054.93  &  5055.98  &  Si II(5)     & 253  & 9  &  -62.2  \\
5126.24  &  5127.35  &  Fe III(5)    & 68   & 8  &  -64.8  \\
5157.74  &  5158.75  &  Fe III(5)    & 159  & 11 &  -58.6  \\
5196.83  &  5197.90  &  [N I](1F)    & 111  & 9  &  -61.7  \\
5199.26  &  5200.26  &  [N I](1F)    & 53   & 6  &  -57.6  \\
5242.02  &  5243.31  &  Fe III       & 46   & 4  &  -73.7  \\
5260.56  &  5261.61  &  [Fe II](19F) & 128  & 8  &  -59.8  \\
5272.22  &  5273.35  &  [Fe II](18F) & 75   & 5  &  -64.2  \\
5297.98  &  5298.97  &  O I(26)      & 86   & 7  &  -56.0  \\
5511.65  &  5512.77  &  O I(25)      & 54   & 5  &  -60.9  \\
5553.87  &  5554.95  &  O I(24)      & 65   & 6  &  -58.2  \\
5753.44  &  5754.59  &  [N II](2D)   & 144  & 8  &  -59.9  \\
5956.34  &  5957.56  &  Si II(4)     & 104  & 5  &  -61.4  \\
5957.30  &  5958.58  &  O I          & 118  & 6  &  -64.4  \\
5977.72  &  5978.93  &  Si II(4)     & 239  & 7  &  -60.6  \\
6031.08  &  6032.67  &  Fe III       & 58   & 8  &  -79.0  \\
6045.17  &  6046.38  &  O I(22)      & 242  & 8  &  -59.9  \\
6299.04  &  6300.30  &  [O I](1F)    & 431  & 7  &  -59.9  \\
6345.77  &  6347.11  &  Si II(2)     & 381  & 7  &  -63.2  \\
6362.55  &  6363.78  &  [O I](1F     & 126  & 6  &  -57.9  \\
6370.05  &  6371.37  &  Si II(2)     & 162  & 5  &  -62.1  \\
6546.75  &  6548.05  &  [N II](1F)   & 5270 & 25 &  -59.5  \\
6561.49  &  6562.82  &  H I          & 44412 & 178 & -60.6  \\
6582.13  &  6583.45  &  [N II](1F)   & 16070 & 70 & -60.1   \\
6715.12  &  6716.44  &  [S II](2F)   & 1042 & 10 &  -58.9  \\
6729.48  &  6730.82  &  [S II](2F)   & 2200 & 14 &  -59.6  \\
7000.71  &  7002.13  &  O I(21)      & 317  & 6  &  -60.7  \\
7153.66  &  7155.17  &  [Fe II]      & 101  & 18  & -63.2   \\
7229.69  &  7231.03  &  C II(3)      & 233  & 7  &  -55.5  \\
7234.84  &  7236.02  &  C II(3)      & 317  & 5  &  -48.8  \\
7252.96  &  7254.36  &  O I(20)      & 292  & 9  &  -57.8  \\
7317.53  &  7318.99  &  [O II]       & 105  & 35 &  -59.8  \\
7318.64  &  7319.99  &  [O II]       & 960  & 8  &  -55.2  \\
7328.23  &  7329.67  &  [O II]       & 510  & 8  &  -58.8  \\
7329.29  &  7330.73  &  [O II]       & 500  & 15  & -58.8   \\
7376.42  &  7377.83  &  [Ni II](2F)  & 372  & 18  & -57.3   \\
7410.16  &  7411.61  &  [Ni II](2F)  & 122  & 14  & -58.6   \\
7466.81  &  7468.31  &  N I(3)       & 215  & 7   & -60.2   \\
7894.45  &  7896.37  &  Mg II(8)     & 78   & 7   & -72.8   \\
7998.29  &  8000.07  &  [Cr II](1F)  & 88   & 6   & -66.6   \\
8123.62  &  8125.30  &  [Cr II](1F)  & 52   & 6   & -61.9   \\
8240.66  &  8242.34  &  N I(7)       & 291  & 14  & -61.0   \\
8299.81  &  8301.59  &  [Ni II](2F)  & 250  & 40  & -64.2   \\
8444.79  &  8446.25  &  O I(3)       & 6090 & 19  & -51.8   \\
8701.50  &  8703.25  &  N I(1)       & 221  & 15  & -60.2   \\
8709.93  &  8711.70  &  N I(1)       & 268  & 16  & -60.9   \\
9067.12  &  9068.80  &  [S III]      & 1629 & 14  & -55.5   \\

\hline

\end{longtable}
\end{center}

\onecolumn
\onecolumn
\section{ Radial velocities  and elemental abundances of the stellar absorption lines} \label{vr}

\footnotesize
\begin{center}
\begin{longtable}{cccccccccc}
  \caption{Absorption lines used for measurement of the stellar radial velocity and elemental abundances.}
\\

\hline
  Ion & $\lambda_{lab}$, \AA & $\lambda_{obs}$, \AA & $E_l$, eV & EW, m\AA & $\sigma_{EW}$, m\AA & $\log \epsilon$ & $\sigma_{\log \epsilon}$ & $V_r$, km s$^{-1}$ & $\sigma_{V_r}$, km s$^{-1}$ \\
      \hline

  \endfirsthead

   \multicolumn{2}{l}{continued Table\ref{vr}}\\
  Ion & $\lambda_{lab}$, \AA & $\lambda_{obs}$, \AA & $E_l$, eV & EW, m\AA & $\sigma_{EW}$, m\AA & $\log \epsilon$ & $\sigma_{\log \epsilon}$ & $V_r$, km s$^{-1}$ & $\sigma_{V_r}$, km s$^{-1}$ \\

   \hline
   \endhead
        N II  &  3995.00  &  3994.29  &  18.5  &  135.4  &  15.0  &  7.48  &  0.04  & -53.0 & 2.3 \\
        N II  &  5001.48  &  --       &  20.7  &   38.0  &  10.3  &  7.20  &  0.05  & --    & --  \\
        N II  &  5045.10  &  --       &  18.5  &   51.6  &  9.1   &  7.90  &  0.03  & --    & --  \\
        N II  &  5666.62  &  5665.79  &  18.5  &   50.2  &  4.5   &  7.20  &  0.03  & -44.3 & 1.6 \\
        N II  &  5676.02  &  5675.13  &  18.5  &   52.1  &  5.0   &  7.56  &  0.02  & -46.9 & 1.7 \\
        N II  &  5679.59  &  5678.80  &  18.5  &   92.4  &  5.0   &  7.77  &  0.03  & -40.1 & 1.2 \\
        \hline

        O II  &  3911.97  &  3911.21  &  25.7  &  136.8  &  12.0  &  8.59  &  0.04  & -58.3 & 2.7 \\
        O II  &  3954.36  &  --       &  23.4  &  165.9  &  16.0  &  8.74  &  0.05  &  --   & --  \\
        O II  &  3982.71  &  3982.12  &  23.4  &  100.6  &  10.0  &        &        & -44.6 & 4.0 \\
        O II  &  4069.78  &  4069.04  &  25.6  &  209.7  &  15.0  &  8.51  &  0.02  & -54.3 & 1.3 \\
        O II  &  4072.14  &  4071.44  &  25.6  &  185.7  &  10.0  &  8.59  &  0.02  & -51.5 & 1.3 \\
        O II  &  4132.80  &  --       &  25.8  &   94.0  &  11.0  &  8.57  &  0.05  &  --   & --  \\
        O II  &  4153.29  &  4152.45  &  25.8  &  139.0  &  10.0  &        &        & -60.7 & 1.6 \\
        O II  &  4317.15  &  4316.39  &  23.0  &  160.0  &   8.0  &        &        & -53.1 & 1.3 \\
        O II  &  4319.63  &  4318.91  &  23.0  &  195.8  &  16.0  &  8.56  &  0.01  & -50.0 & 1.0 \\
        O II  &  4345.52  &  4344.77  &  23.0  &  199.0  &   8.0  &  8.61  &  0.02  & -51.4 & 1.0 \\
        O II  &  4347.44  &  4346.58  &  25.7  &  160.0  &  21.0  &  8.52  &  0.01  & -59.1 & 1.3 \\
        O II  &  4349.40  &  4348.69  &  23.0  &  343.0  &  10.0  &  8.52  &  0.01  & -48.8 & 0.6 \\
        O II  &  4351.27  &  4350.50  &  25.7  &  182.5  &   8.0  &  8.55  &  0.02  & -53.2 & 0.9 \\
        O II  &  4366.89  &  --       &  28.9  &  192.3  &  16.0  &  8.52  &  0.03  & --    & --  \\
        O II  &  4414.90  &  4414.15  &  23.4  &  389.0  &  31.0  &        &        & -50.8 & 0.8 \\
        O II  &  4416.97  &  4416.23  &  23.4  &  319.0  &  26.0  &  8.51  &  0.03  & -50.3 & 1.1 \\
        O II  &  4590.97  &  4590.12  &  25.7  &  193.0  &   7.0  &  8.63  &  0.02  & -55.3 & 0.9 \\
        O II  &  4596.16  &  4595.28  &  25.7  &  155.0  &   7.0  &  8.68  &  0.02  & -57.6 & 1.2 \\
        O II  &  4638.85  &  4638.05  &  23.0  &  236.6  &   6.0  &  8.65  &  0.02  & -51.4 & 0.7 \\
        O II  &  4641.81  &  4641.03  &  23.0  &  312.1  &   6.0  &  8.54  &  0.01  & -50.4 & 0.5 \\
        O II  &  4661.63  &  4660.85  &  23.0  &  236.6  &   6.0  &  8.61  &  0.02  & -50.2 & 0.7 \\
        O II  &  4673.73  &  --       &  23.0  &   46.0  &   5.0  &  8.53  &  0.02  & --    & --  \\
        O II  &  4676.24  &  4675.51  &  23.0  &  183.6  &   7.0  &  8.61  &  0.02  & -46.8 & 1.0 \\
        O II  &  4699.16  &  4698.33  &  28.5  &   80.7  &   5.0  &        &        & -53.0 & 1.5 \\
        O II  &  4703.16  &  --       &  28.5  &   41.1  &   8.0  &  8.52  &  0.02  & --    & --  \\
        O II  &  4705.35  &  --       &  28.2  &  125.0  &   9.0  &  8.49  &  0.02  & --    & --  \\
        O II  &  4710.01  &  --       &  26.2  &   61.9  &   9.0  &  8.69  &  0.03  & --    & --  \\
        O II  &  4906.83  &  --       &  26.3  &   45.4  &   8.0  &  8.49  &  0.04  & --    & --  \\
        \hline

        Mg II  & 4481.23  &  --       &  29.6  &   51.7  &   7.0  &  6.56  &  0.04  & --    & --  \\
        \hline

        Al III & 5696.60  &  --       &  15.6  &   43.8  &   5.0  &  5.25  &  0.05  & --    & --  \\
        \hline

        Si IV  &  4116.10 &  --       &  24.1   &  127.5 &   7.0  &  7.00  &  0.04  & --    & --  \\
        Si III &  4567.84 &  --       &  19.0   &  288.8  &  7.0  &  7.00  &  0.01  & --    & --  \\
        Si III &  4574.76 &  --       &  19.0   &  170.7  &  7.0  &  7.14  &  0.01  & --    & --  \\
        Si III &  5739.73 &  5738.70  &  19.7   &  172.0  &  4.0  &  7.02  &  0.01  & -54.4 & 0.8 \\
        \hline

        S III  &  4253.59 &  --       &  18.2   &  120.3  &  10.0 &  6.29  &  0.03  & --    & --  \\
        S III  &  4284.97 &  --       &  18.2   &   82.6  &  13.0 &  6.33  &  0.04  & --    & --  \\
        \hline
        C II   &  3918.97 &  --       &  16.3   &  119.5  &  34.0 &  7.73  &  0.21  & --    & --  \\
        C III  &  4647.42 &  --       &  29.5   &   82.3  &  16.0 &  7.84  &  0.19  & --    & --  \\
        \hline

\end{longtable}
\end{center}

\end{document}